\documentclass[aip,jcp,reprint]{revtex4-1}
\usepackage{amsmath}
\usepackage{amssymb}
\usepackage{graphicx}
\usepackage{time}
\usepackage{siunitx}
\usepackage{pbox}
\usepackage{bm} 
\usepackage{subfig}
\usepackage{textgreek}
\usepackage[colorlinks,linkcolor=blue,citecolor=blue,urlcolor=blue]{hyperref}
\usepackage{float}
\graphicspath{ {./} }
\begin{document}

\newcommand{\average}[1]{\left\langle {#1} \right\rangle} 
\newcommand{\pd}[2]{\frac{\partial {#1}}{\partial {#2}}}
\newcommand{\cov}{\textrm{Cov}}
\newcommand{\var}{\textrm{Var}}

\title{Using reweighting and free energy surface interpolation to predict solid-solid phase diagrams}

\author{Natalie P. Schieber}
\affiliation{Department of Chemical and Biological Engineering, University of Colorado Boulder, Boulder, CO 80309, USA}
\author{Eric C. Dybeck}
\affiliation{Department of Chemical Engineering, University of Virginia, Charlottesville, VA 22904, USA}
\author{Michael R. Shirts}
\affiliation{Department of Chemical and Biological Engineering, University of Colorado Boulder, Boulder, CO 80309, USA}

%%%%%%%%%%%%%%%%%%%%%%%%%%%%%%%%%%%%
%	Abstract 	           %
%%%%%%%%%%%%%%%%%%%%%%%%%%%%%%%%%%%%
\begin{abstract}
Many physical properties of small organic molecules are dependent on the current crystal packing, or polymorph, of the material, including bioavailability of pharmaceuticals, optical properties of dyes, and charge transport properties of semiconductors. Predicting the most stable crystalline form requires determining the crystalline form with the lowest relative Gibbs free energy. Effective computational prediction of the most stable polymorph could save significant time and effort  in the design of novel molecular crystalline solids or predict their behavior under new conditions.

In this study, we introduce a new approach using multistate reweighting to address the problem of determining solid-solid phase diagrams, and apply this approach to the phase diagram of solid benzene.
For this approach, we perform sampling at a selection of temperature and pressure states in the region of interest. We use multistate reweighting methods to determine the reduced free energy differences between $T$ and $P$ states within a given polymorph. The relative stability of the polymorphs at the sampled states can be successively interpolated from these points to create the phase diagram by combining these reduced free energy differences with a reference Gibbs free energy difference between polymorphs. The method also allows for straightforward estimation of uncertainties in the phase boundary. We also find that when properly implemented, multistate reweighting for phase diagram determination scales better with size of system than previously estimated.
\end{abstract}
\maketitle
%%%%%%%%%%%%%%%%%%%%%%%%%%%%%%%%%%%%
\newpage
\section{Introduction}

The overall packing of a crystalline compound has a large effect on the properties and applications of the material. Polymorphism is the ability of a molecule to exist in more than one crystalline configuration, or polymorph. Physical and chemical properties of the same substance in different polymorphic forms are not guaranteed to be the same. Therefore the polymorphic form of a material affects its utility at ambient conditions, the polymorph present can determine the sensitivity to detonation~\cite{Eckhardt2007, Fabbiani2006}. Polymorphism has also been shown to affect the strength properties of concrete~\cite{Stanek2002} and charge transport properties in semiconductor materials~\cite{Stevens2015}. 

One of the most critically important areas where prediction of polymorphism is important is in 
pharmaceutical formulation.
In multiple instances previously unknown polymorphic forms of solid state drugs resulted in disruptions in market availability or patent litigation. Many of these cases result from the recrystallization of a material into a different polymorph during or after production. This occurs when the manufactured polymorph is not the globally most stable structure under ambient conditions, and the material eventually recrystallizes into the more stable structure.

This latent polymorphism has affected the market availability of pharmaceuticals. Since patents are typically issued for a particular crystalline structure of a pharmaceutical, knowledge of the most stable structure is important to protect intellectual property ~\cite{Bucar2015, Racokzy2005}.  In 2003 GlaxoSmithKline lost a court case in which a generic firm began making an off-patent polymorph of a patented drug ~\cite{Racokzy2005}. 
Recrystallization into more stable polymorphs has also led to market disruptions and recalls. Two examples of this are the pharmaceuticals Rotigotine and Ritonavir ~\cite{Chen2009, Bauer2001a}. 

Pressure dependence on polymorphism is important in the production processes of pharmaceuticals. During the production of many drugs, the materials undergo processes such as milling and tabletting, which expose the crystal to high pressures for short periods of time. These pressures can affect the stability of various polymorphs. In one study, of 32 drugs studied, 11 were shown to have the potential for polymorphism at the pressures used in milling processes~\cite{Boldyrev2004}. One specific example, is the antimicrobial drug phenylbutazone. This compound exists in three forms ($\alpha$, $\beta$, and $\delta$), at room temperature. After grinding, another form, $\epsilon$, was found to be the predominant occurring form~\cite{Fabbiani2009}.  For these reasons, it is important to know not only the dependence of polymorph stability on temperature, but also on pressure. 

Full temperature and pressure phase diagrams are also important in the fields of geophysics and astrophysics. Materials present in places such as asteroids, or the mantle and core of the Earth, are subject to extreme temperatures and pressures. The full temperature-pressure phase diagram of iron at pressures up to 200 GPa and temperatures up to 4500K was determined experimentally by Boehler et al.~\cite{Boehler1985} and a potential new polymorph was found. In another case, Choukroun et al. determined the phase diagram of the ammonia-water system, which has shown to be important in the study of nebula formation~\cite{Choukroun2010}.

Predicting polymorph stability experimentally is expensive, and has the potential to miss polymorphs. Experimental determination of the structure of synthesized polymorphs relies on methods such as x-ray scattering and Raman spectroscopy~\cite{Thiery1988, Aaltonen2003}. The polymorph obtained on the initial synthesis is not guaranteed to be the globally most stable, and in experimental testing, polymorph stability must be determined at one $T$,$P$ point at a time instead of generating the entire diagram at once. Computational modeling for phase diagram prediction has the potential to be  
a cheaper and more efficient alternative for systems where models are sufficiently accurate and efficient. Even if not perfect, computational studies can help to guide experimental studies and point out polymorphs that may not be caught experimentally. For example, new polymorphs of 5-fluorouracil and aspirin were found experimentally after being predicted computationally  \cite{Price2005z, Vishweshwar2005}.

This project is motivated by the need for an improved solid state phase diagram prediction method that takes into account both temperature and pressure. Such a method should be able to determine the relative thermodynamic stability of different polymorphs of a material at a range of temperatures and pressures. Knowing the most stable polymorph at each temperature and pressure can ensure that no latent polymorphism, or recrystallization to previously unknown polymorphs, is observed and that the storage temperature and pressure of the drug will be correct to avoid phase transitions. Accurate phase diagrams allow for the storage temperature and processing methods to be chosen to avoid recrystallization.~\cite{Price2004, Dunitz2003} 

Previous methods for phase diagram prediction have limitations for solid-solid phase coexistence, making development of a novel phase diagram approach for small molecules desirable. 
There are a variety of suitable methods for the prediction of fluid-fluid coexistence, but methods for systems including solids systems are still frequently inadequate. 
In this project, we focus on improvements to the calculation of phase diagrams specifically for solid-solid systems, 
though the approach is likely to be useful for solid-liquid systems.
Other previous methods exist for the calculation of vapor-liquid equilibria such as the group contribution concept~\cite{Yan1999}, integral equation theory~\cite{Cheng1993}, Gibbs ensemble technique~\cite{Panagiotopoulos1995a}, or Gibbs-Duhem integration~\cite{Kofke1993, Strachan1999}. 

\begin{center}
\begin{figure}
        \includegraphics[width=0.35\textwidth]{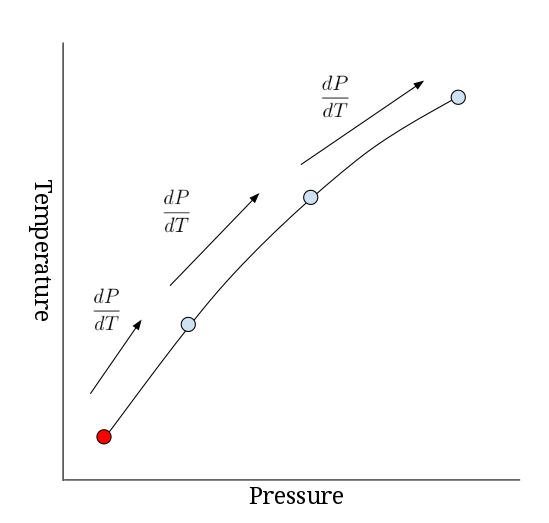} 
        \caption{\label{fig:gdit} Gibbs-Duhem integration uses a known coexistence point at a start for integration along the coexistence line.}
\end{figure}
\end{center}

The Gibbs ensemble technique~\cite{Panagiotopoulos1995a, Panagiotopoulos1988, Panagiotopoulos1995, Panagiotopoulos2002} is a phase diagram prediction method that is useful for vapor-liquid coexistence. It uses the equilibration of volume and chemical potential between two simulation volumes to determine the equilibrium pressure at a specified temperature. Two simulation volumes are run in parallel. The initial conditions are the temperature of the desired coexistence point and an estimated pressure. As the simulation progresses, Monte Carlo moves are performed to equilibrate the pressure, volume and chemical potential.
There are both advantages and disadvantages to the Gibbs ensemble technique. It does not require any coexistence points to be known a priori and is useful for systems of lower densities. However, it does require that the initial pressure be close enough to the coexistence pressure that volume emptying, where a starting point too far from equilibrium causes Monte Carlo steps that move all molecules to one simulation volume, is not observed in one of the volumes~\cite{VantHof2006}. It is also not useful for solid crystalline systems because of the particle insertion Monte Carlo step, which is not favorable in crystalline systems. 
\begin{equation} \label{eq:clausius}
\frac{dP}{dT}_{\mathrm{phase-equilibrium}} = -\frac{\Delta H}{T \Delta V}
\end{equation}
Another commonly used phase diagram prediction method is Gibbs-Duhem integration. This method relies on the Clausius-Clapeyron relationship to provide a differential equation for relating the change in equilibrium pressure to the change in equilibrium temperature~\cite{Kofke1993, Strachan1999}. A point on the coexistence line between phases is required to start Gibbs-Duhem integration. Simulations are performed in both phases at that point to determine the difference in enthalpy and molar volume between the phases. Eq.~\ref{eq:clausius} and numerical integration are then used to determine the next point on the coexistence line. At each step, predictor-corrector equations are used to solve for successive points along the phase coexistence line.
There are a range of numerical integration techniques that can be used, giving a range of tradeoffs in accuracy, stability, and efficiency~\cite{Kofke1993}. 
This procedure is repeated until the desired coexistence line has been built. This process can be seen in Figure \ref{fig:gdit}. The sources of error in this method include the accuracy of the initial coexistence point, the integration method used, the temperature step size~\cite{Kofke1993} and the distance from the initial coexistence point~\cite{VantHof2006}. 
A variant of Gibbs-Duhem integration is ``advanced Gibbs-Duhem integration''~\cite{VantHof2006}, introduced by Van 't Hof et al. In this case, Gibbs-Duhem integration is supplemented by multiple-histogram reweighting at nearby states.  A step of Gibbs-Duhem integration is carried out, and some number of simulations close in state space are chosen nearby. Multiple-histogram reweighting~\cite{Ferrenberg1989, Kumar1992} is used to combine these additional simulations and compute the terms involved in the Clapeyron equation more accurately than in the initial pass. The expectations required, such as enthalpy and volume, can be estimated at any value of $T$ and $P$ by reweighting, allowing the integration to be carried out with accuracy limited only by the statistical errors of the reweighting process. Like the original Gibbs-Duhem integration, this method still accrues error by virtue of being a numerical integration and requires a priori knowledge of a coexistence point, which also contributes error. This method also requires a sufficiently low free energy barrier between phases and histogram overlap between the phases. 

Histogram reweighting approaches have also been previously used to compute phase equilibrium lines in conjunction with reservoir grand canonical Monte Carlo and growth expanded ensemble. The method of Rane et al. \cite{Rane2013} starts with a phase coexistence point and uses grand canonical and isothermal-isobaric temperature-expanded ensemble (TEE) methods which have subensembles that differ in temperature. These TEE ensembles are then run using a variety of temperatures to determine where the free energy of the phases is equal. Multiple histogram reweighting methods are used to refine the initial prediction. This approach has been used in computing phase behavior in diamond and simple cubic lattice structures \cite{Jain2013} and the critical point of mixtures of molecular fluids \cite{Tamaghna2017}. The use of growth expanded ensemble overcomes the difficulties in the insertion Monte Carlo step of Gibbs ensemble. This method uses an initial coexistence point that is not a phase equilibria point, bit requires the use of expanded ensemble Monte Carlo methods, which limits the simulation packages that can be used.

Another existing phase diagram prediction method, free energy extrapolation~\cite{Escobedo2014}, finds the slope of the free energy surface and extrapolates it to nearby points to find coexistence points. In this method, the probability of a configuration in a simulation having a certain volume and energy is assumed to be a Gaussian. The slopes of the free energy surfaces are first determined at a reference point, where the difference in free energy is known. A fit is then used to extrapolate this slope to nearby points. From this estimated slope, and the reference difference in free energy, the difference in coexistence in the $f_1$ and $f_2$ values is found by using Eq.~\ref{eq:fenex}, where $\Delta f_1$ is the size of the integration step to be used. In this equation, $\phi_{i}^{2} - \phi_{i}^{1}$ is the difference in free energy at the reference point, $f_1$ and $f_2$ are the dimensions along which coexistence is being studied (for example, temperature and pressure), and $\cov^{I}_{12}$ is the covariance in phase $I$ between dimensions 1 and 2. This method is applicable to any two thermodynamic dimensions, but $f_1$ and $f_2$ would typically be temperature and pressure. Starting with a point of either known coexistence or known free energy difference, simulations are performed at point $i$ in both phases to determine all of the required values. The next coexistence point is then found, and the process is repeated until the entire line has been found by integration. This method requires serial simulations along the line, and the uncertainty accumulates along the line.
\begin{widetext}
\begin{equation} \label{eq:fenex}
\Delta f_{2} = \frac{ \phi_{i}^{2} - \phi_{i}^{1} + \Delta f_{1} ( \bar{x}_{1,i}^{2} - \bar{x}_{1,i}^{1} ) - \frac{1}{2} (\cov_{11,i}^{2} - \cov_{11,i}^{1} ) (\Delta f_1)^{2} }{ -(\bar{x}_{2,i}^{2} - \bar{x}^{1}_{2,i}) + (\cov^{2}_{12,i} - \cov^{1}_{12,i}) \Delta f_{1}  +\frac{1}{2} \Delta f_{2} (\cov^{2}_{22,i} - \cov^{1}_{22,i}) }
\end{equation} 
\end{widetext}

The methods above, with the exception of free energy extrapolation, require initial coexistence points, which can be difficult to obtain.  There are multiple ways of obtaining a coexistence point, but all have complications~\cite{Zhang2012}. One standard method uses Gibbs ensemble simulations. A single temperature run of the Gibbs ensemble method described previously will provide an initial coexistence point. However, this suffers from the same previously described drawbacks of the Gibbs ensemble for solid simulations~\cite{Panagiotopoulos1995}.  A coexistence point could in principle be found by direct simulation along a thermodynamic variable, for example, where a single simulation is performed in increasing temperature steps until phase change is observed~\cite{Broughton1987}. This often results in hysteresis in the phase change temperature due to the thermodynamic barrier between states~\cite{Strachan1999,Agrawal2003} and the phase change point in one direction does not match the point when going in the other direction, introducing inaccuracy. Similarly, voids in the crystal can be added and the apparent melting point measured as a function of void fraction until it levels off, which is the melting point~\cite{Agrawal2003}. Another way of finding coexistence is to run the pseudo-supercritical path (PSCP) method of Eike et al. at temperatures near the expected melting point~\cite{Eike2006}. Gibbs Duhem integration is then used to find the coexistence point, where the free energy found by the PSCP is 0~\cite{Zhang2012, Eike2006}.

We present a new approach to phase coexistence prediction, the Successive Interpolation of Multistate Reweighting (SIMR) method, aimed at solid state systems but which should be applicable for any condensed phases. It borrows many of the ideas from previous methods, but also overcomes many of the issues raised by these methods. This method uses the Gibbs free energy difference between phases to provide both the coexistence lines and a quantitative measure of relative stability throughout the entire region studied,  indicating regions where the free energy difference is small, 
and showing the general trends of how the stability changes with temperature. This method relies on multistate reweighting, a statistical mechanical method that using importance sampling to take information from the Boltzmann distribution of sampled states to extrapolate to nearby states \cite{Shirts2008}. Multistate reweighting is the binless version of the multiple-histogram reweighting technique discussed earlier~\cite{Ferrenberg1989}, with improved accuracy and much simpler interpretation and calculation~\cite{shirts_mbar_arxiv_2017}.
Because all simulations can be run independently this approach allows simulations run in parallel to improve wall clock computational time. It allows the easy calculation of uncertainty, which is not propagated and is therefore not a function of distance from the reference point. We demonstrate SIMR by calculating phase diagrams of solid benzene 
calculated using full molecular dynamics.

SIMR uses local reconstruction of the free energy surfaces $G(T,P)$ of each polymorph. If we know the difference $G_i(T,P)-G_j(T,P)$ between any two polymorphs at any given states, we can find where the two surfaces lie with respect to each other, and identify the most thermodynamically stable structure at any temperature and pressure. The coexistence lines between the polymorphs are then the line of intersection between the $G(T,P)$ surfaces of any two polymorphs. For the SIMR method specifically, a combination of reweighting methods is used to obtain the Gibbs free energy difference at each point. These reweighting methods estimate the free energy difference between thermodynamic states using the probability distribution at each of multiple states. 

\section{Theory}

\subsection{Multistate Reweighting}
The SIMR method uses multistate reweighting as implemented in the multistate Bennett acceptance ratio (MBAR) to calculate free energy differences between temperature and pressure points within a polymorph. MBAR estimates the reduced free energies $f_i$ of all states of interest $i$ by solving the system of nonlinear equations, for each state's reduced free energy, $f_{i}$ relative to the free energies of the other states, $f_{k}$,\cite{Shirts2008} as shown in Eq.~\ref{eq:mbar}. 
			\begin{equation} \label{eq:mbar}
				f_{i} = -  \ln \sum_{j=1}^{K} \sum_{n=1}^{N_{j}} \frac{\exp[- u_{i}(x_{jn})]}{\sum_{k=1}^{K} N_{k} \exp[ f_{k} - u_{k}(x_{jn})]}
			\end{equation}
MBAR has been proven to be the statistically most efficient estimator of thermodynamics properties with more than two thermodynamic states~\cite{Shirts2008, Chodera2007}.  Reweighting at a number of different $T$ and $P$ points makes it possible to easily estimate the reduced free energy differences between temperature and pressure states within a polymorph, but not the differences between polymorphs. MBAR has been implemented as a Python package, \texttt{pymbar} version 3.0.0 (\url{http://www.github.com/choderalab/pymbar}), which was used for all calculations. 

In the constant pressure and temperature (NPT) ensemble, $u_i(x_{nj}) = \beta_i U(x_{nj}) + \beta_i P_i V(x_{nj})$ where $u_i(x_{jn})$ is the reduced energy of configuration $x_{nj}$ sampled from state $j$, evaluated in state $i$. This results in a reduced free energy that is related to the Gibbs free energy by $ f_{k} = \beta G_{k}$. In the constant volume and temperature, NVT ensemble, the reduced free energy is simply $ u_i(x_{nj}) = \beta_i U(x_{nj}) $ and the reduced free energy is then related to the Helmholtz free energy as $ f_{k} = \beta_k A_{k}$. If the states of interest differ only by temperature and pressure, then $u_i(x)$ and $u_j(x)$ differ only by $\beta$ and $P$, and thus recalculating the reduced energy at each state can be done entirely in postprocessing if the total energy and volume are saved for each uncorrelated configuration $x$. This avoids having to re-evaluate the potential energies of the configurations in a new potential, as is typically needed for alchemical reweighting approaches, where $U(x)$ changes between states. 

\subsection{Pseudo-supercritical Path}

\begin{figure*}
 \includegraphics[width=0.8\textwidth]{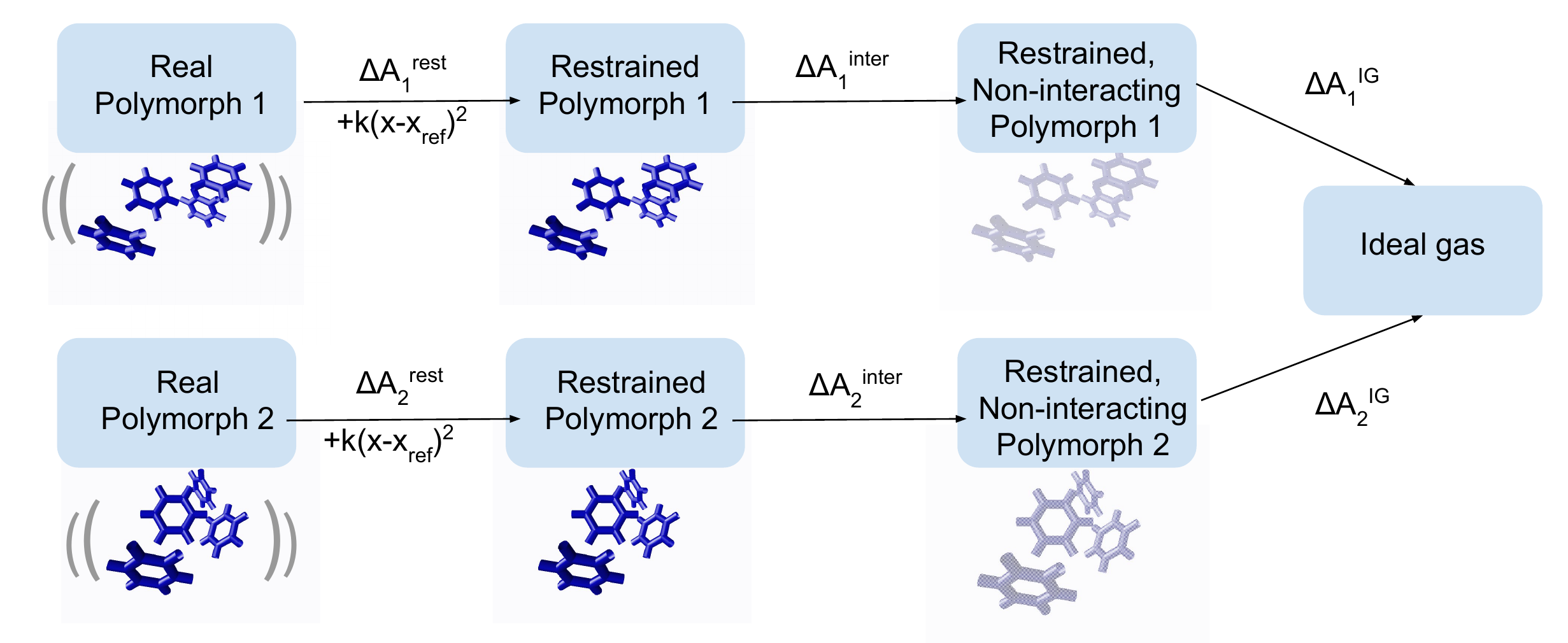} 
        \caption{The PSCP process uses three steps per polymorph to calculate the free energy between a real crystal and an ideal gas and thus the free energy difference between polymorphs. Adapted from Dybeck et al.~\cite{Dybeck2016}\label{fig:pscpdybeck}}
\end{figure*}

In order to obtain a $\Delta G$ value between each set of polymorphs, the reduced free energy values within polymorphs obtained by MBAR must be combined with a reference Gibbs free energy difference between polymorphs at the same $T$ and $P$. While the reference free energy can be obtained using a variety of methods, such as metadynamics~\cite{Laio2008, Barducci2008} and the Frenkel-Ladd method~\cite{Frenkel1984}, here the reference free energy difference is calculated using a pseudo-supercritical path (PSCP)~\cite{Zhang2012, Eike2006, Eike2005}. 
The reference free energy must be at a point where the less-stable phase is kinetically trapped, which can generally be a moderate distance from the phase equilibrium line for solid-solid equilibria, and often for liquid-solid equilibria as well. 
The PSCP creates a closed thermodynamic cycle in which the two polymorphs to be compared are brought from a real crystal to an ideal gas. In the ideal gas state, the Helmholtz free energy between all polymorphs is zero, so by calculating the free energy to bring the crystal from physical crystal to ideal gas, the difference between the real crystal polymorphs can be found. Multistate reweighting with MBAR is used to calculate the free energy differences along the thermodynamic path. 
The specific details for this procedure have been described in Dybeck
et al.~\cite{Dybeck2016}, but a schematic of this process is shown in
Fig.~\ref{fig:pscpdybeck}.

Briefly, the PSCP for computing the free energy between two solid polymorphs is constructed by summing three steps for each polymorph. The first step is to atomically restrain the polymorphs to near their equilibrium positions. This is done using a $ \lambda_{rest} $ value, which is a coupling parameter representing the strength of the restraints imposed on the molecules. Simulations are performed at twenty values of a harmonic restraint from $ \lambda_{rest} = 0 $ which is unrestrained to $ \lambda_{rest} = 1 $ which is fully restrained, spaced quadratically with respect to the spring constant. Multistate reweighting using the twenty states of varying $\lambda $ parameters is then used to find the free energy difference between the $\lambda=0$ and $\lambda=1$ states, which is then used to find the $ \Delta A_{rest} $ value for the polymorph.   A range of different paths can be used, and if correctly implemented, will differ only in their efficiency. 

The second step in the PSCP is to remove the intermolecular interactions between molecules while leaving the intramolecular interactions. This step uses another coupling parameter, $ \lambda_{inter} = 0 $, which scales the amount of the intermolecular potential energy included in the Hamiltonian, shown in Eq.~\ref{eq:lambdaint}, where  $\eta$ is a scaling parameter chosen for the system between 0 and 1, $U_{i} $ is the potential used, and  $ U_{inter}$ is the raw intermolecular potential~\cite{Eike2005}. At this step the lambda value for the restraints is 1. Simulations were performed at ten quadratically spaced values from  $ \lambda_{inter} = 1 $, which is fully interacting, to $ \lambda_{inter} = 0 $, which is non-interacting.  Multistate reweighting is then used with the ten states of varying $\lambda_{rest}$ values  to find the free energy difference associated with removing intermolecular interactions, and thus the $ \Delta A_{inter} $ value for the polymorph. The third step is to remove the restraints from the non-interacting polymorphs to obtain the ideal gas state. However, this is not necessary because the free energy difference of the restrained non-interacting polymorphs is by definition zero. A schematic of this process is shown in Fig.~\ref{fig:pscpdybeck}. The full equation used to calculate the PSCP value for a single polymorph is given in Eq. 5 of Dybeck et al. \cite{Dybeck2016}.
\begin{equation} \label{eq:lambdaint}
U_{i} = (1-\lambda_{inter} (1- \eta))U_{inter}
\end{equation}
\begin{equation} \label{eq:pscpda}
\begin{split}
\Delta A^{PSCP} = \Delta A^{rest} (\lambda_{rest} = 0 \rightarrow 1) + \\
 \Delta A^{inter} (\lambda_{inter} = 0 \rightarrow 1) + \Delta A_{IG}
\end{split}
\end{equation}

The resulting free energy difference that results from the application of the PSCP to the two crystal polymorphs is the excess Helmholtz free energy $\Delta A^{ex} $ which is converted to the Gibbs free energy by using $ \Delta G = \Delta A^{ex} + \Delta A^{ig} + P \Delta V$, where $  \Delta A^{ig} $ is the ideal gas contribution~\cite{Eike2005}. When performed in the NVT ensemble,  $  \Delta A^{ig} $ is zero. The final equation used to calculate the reference free energy difference for two solid polymorphs is then Eq.~\ref{eq:pscp}. In order to find the free energy difference at a given pressure, the equilibrium volume at that pressure is used in the PSCP calculation. 
\begin{equation} \label{eq:pscp}
\begin{split}
\Delta G_{1\rightarrow 2} = \Delta A_{2,rest}+\Delta A_{2,inter} - \\
\Delta A_{1,rest} - \Delta A_{1,inter} + P \Delta V_{1\rightarrow 2}
\end{split}
\end{equation} 

\subsection{Phase Space Overlap}

When using multistate reweighting to calculate differences in free energy between thermodynamic states, it is essential that there is phase space overlap, or a nonzero probability of a configuration generated from one state (in this case, defined by $T$ and $P$) occurring in another state, in a connected chain of adjacent thermodynamic states connecting the initial and final state of interest. This requirement of configuration space overlap applies whether the difference in thermodynamic states is temperature, pressure, or a value of coupling parameter $\lambda$. This is due to the fact that free energies are essentially ratios of probabilities, and if mutual configurations are not observed between the two states, there can be no way to estimate their free energy difference by statistical mechanics.
This can be observed quantitatively by noting uncertainty estimate in free energy differences using reweighting with two states goes as one over the amount of overlap, as seen in Eq.~\ref{eq:baroverlap}~\cite{Bennett1976}, where $O$ is the overlap and $N_{samples}$ is the number of samples from each state (though the equation is only exact when $N_{samples}$ is equal for both states).  

\begin{figure}
\centering
\subfloat[]{\centerline{\includegraphics[width=0.5\textwidth]{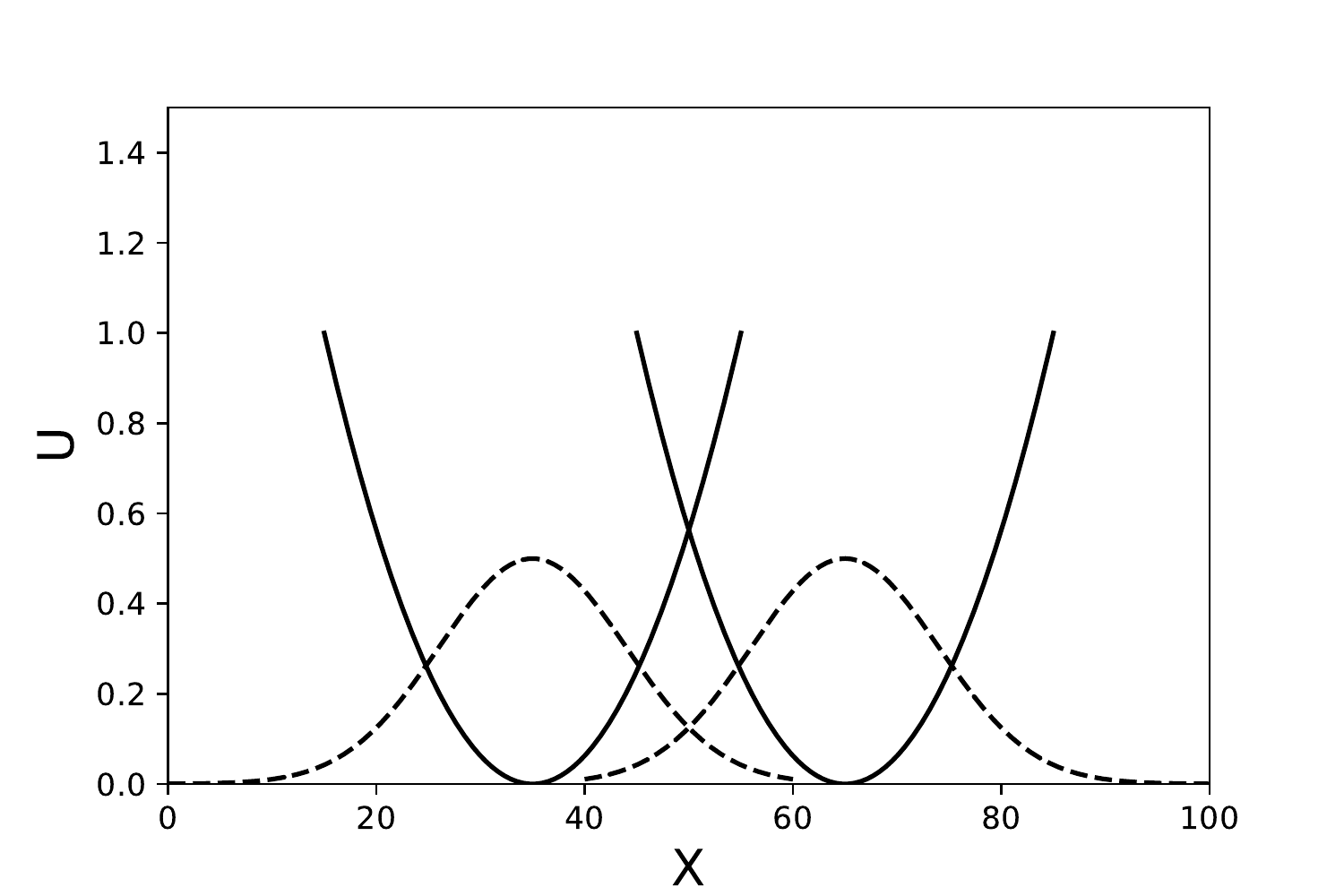}}}
\newline
\subfloat[]{\centerline{\includegraphics[width=0.5\textwidth]{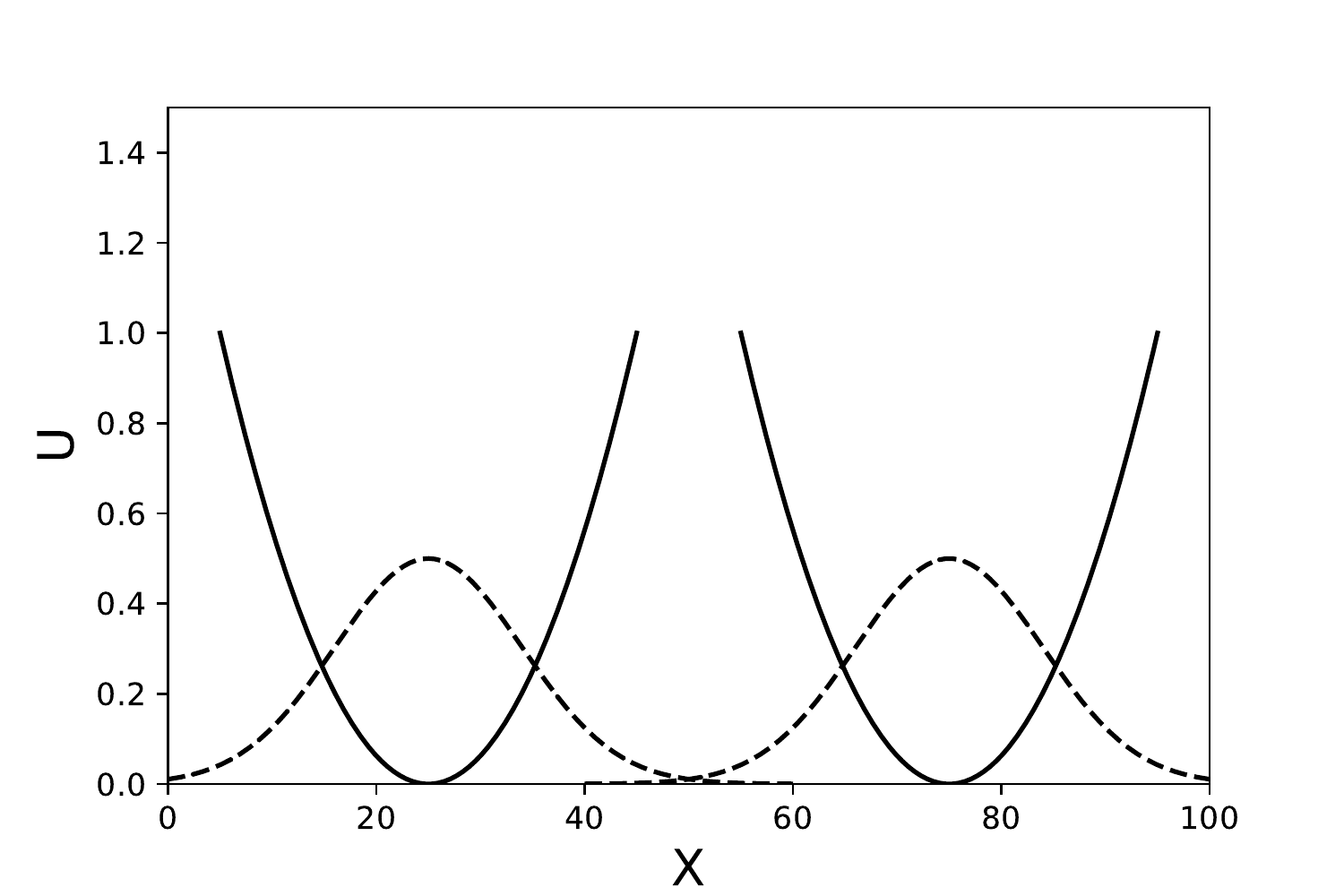}}}
        \caption{\label{fig:overlap1} The potential energy of two harmonic oscillators (solid) and their respective probability distributions in position (dotted). The top oscillators show sufficient phase space overlap for effective free energy difference determination, while the bottom set of oscillators show poor overlap.}
\end{figure}

Assuming a Boltzmann distribution (Eq.~\ref{eq:boltzmann}), where $Z$ is the partition function the overlap $O$ of two distributions over the phase space $\Gamma$ is defined in Eq.~\ref{eq:overlap}~\cite{Zwanzig1954}. This indicates that in order to converge Eq.~\ref{eq:mbar}, the probability of a configuration obtained from state 1 occurring in state 2 must be nonzero. This limits the distance in thermodynamic space that two adjacent states can be placed and still obtain an accurate free energy difference. An example of phase space overlap using harmonic oscillators can be seen in Fig.~\ref{fig:overlap1}. The percent overlap required is dependent on the system and the number of configurations used. For example, in a crystalline benzene test system, the amount of overlap required for \texttt{pymbar} to achieve convergence using 2000 configurations was 0.007 percent but the overlap required to achieve the same uncertainty if only 1000 of those configurations are used would be 0.01. 
\begin{equation} \label{eq:boltzmann}
P(x) = Z^{-1}e^{-\beta U(x)}
\end{equation} 
\begin{equation} \label{eq:baroverlap}
\delta \Delta f = \left(O^{-1}-2\right)^{-1/2} N_{samples}^{-\frac{1}{2}}
\end{equation}
\begin{equation} \label{eq:overlap}
O_{1,2} = \int_{x \in \Gamma} \frac{P_{1}(x) P_{2}(x)}{P_{1}(x)+P_{2}(x)} \,dx
\end{equation} 

\section{Methodology}

We present a proposed new algorithm for the prediction of phase diagrams of small molecules, Successive Interpolation of Multistate Reweighting (SIMR). First, the reference free energy difference is obtained via PSCP.  We perform simulations at varying values of $ \lambda_{rest} $ and $ \lambda_{interact} $, as described previously, in all polymorphs of interest. MBAR, as implemented in \texttt{pymbar}, and a reduced energy definition of $ u_i(x_{nj}) = \beta_i U(x_{nj}) $ are used to determine the reduced free energy between $ \lambda_{rest} = 0 $ and $ \lambda_{rest} = 1 $ for the first leg of the thermodynamic path and $ \lambda_{interact} = 1 $ and $ \lambda_{interact} = 0 $ for the second leg of the path. This is done for all polymorphs and then Eq.~\ref{eq:pscp} is used to obtain the reference Gibbs free energy difference between each set of polymorphs at the specified temperature and the pressure defined by the equilibrium volume. 

Once the reference value has been determined, the first step in the SIMR method is to obtain the free energy differences between temperature and pressure points within each polymorph. Simulations are performed at a set of states in the temperature and pressure of coexistence in all relevant polymorphs. Any set of states with phase space overlap can be chosen, although in this paper, an evenly spaced grid was chosen for simplicity. The reduced energy, $u_i(x_{nj}) = \beta_i U(x_{nj}) + \beta_i P_i V(x_{nj})$ is calculated for each uncorrelated configuration, $x_{n}$, from each state $j$, as evaluated in every other state $i$. These reduced energies are then used with Eq.~\ref{eq:mbar}, as implemented in \texttt{pymbar}, to determine a matrix of reduced free energy differences between every combination of temperature and pressure states within each polymorph. However, because in this case, the temperature is also changing between states, the  $ f_{k}= \beta_k G_{k}$ definition to convert between reduced and Gibbs free energy cannot be used to directly calculate Gibbs free energy differences between states. 

\begin{figure}
        \includegraphics[width=0.5\textwidth]{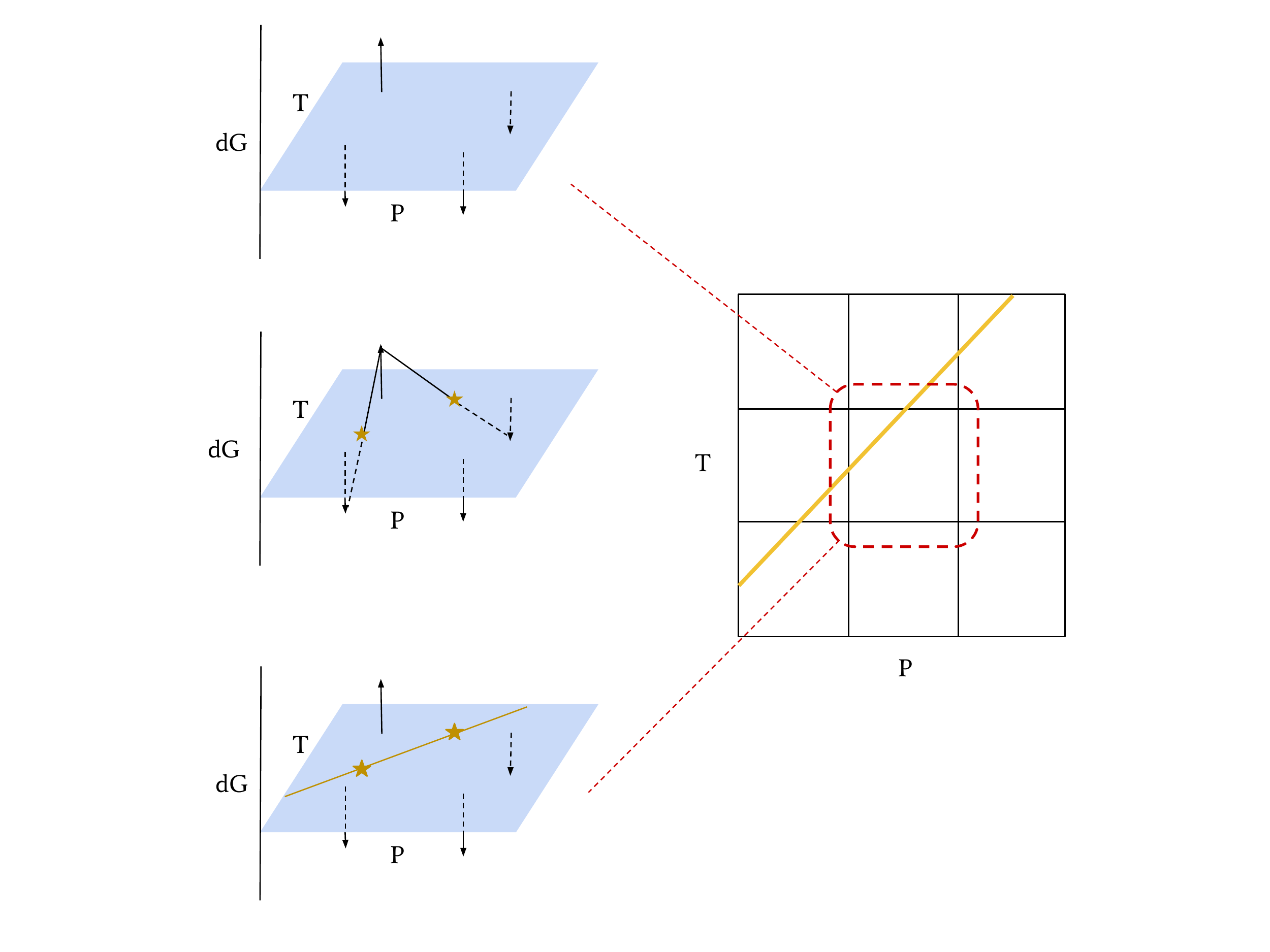} 
        \caption{\label{fig:interp} To predict coexistence points, first the lowest free energy polymorph is determined at each point (top). Then, where the stable polymorph changes between points, the value of the free energy differences are used to find a cross point (middle), and from that a coexistence line is constructed (bottom).}
\end{figure}

To find the Gibbs free energy difference between polymorphs, the reduced free energy differences within polymorphs are then combined with the reference Gibbs free energy between polymorphs obtained from the PSCP~\cite{Dybeck2016}. To do this, the definition of reduced free energy difference between a given point and the reference point, as in Eq.~\ref{eq:singledg}, is used. The difference between two polymorphs is then Eq.~\ref{eq:unreduceddg}, which reduces to Eq.~\ref{eq:finaldg}. This is the final equation used to find the Gibbs free energy difference between polymorphs at each temperature and pressure point in the phase diagram.
		\begin{equation} \label{eq:singledg}
			\beta _{i} G_{i} - \beta_{ref} G_{ref} = f_{i} - f_{ref} 
		\end{equation}
		\begin{equation} \label{eq:unreduceddg}
		\begin{split}
			\beta_{i,1} G_{i,1} - \beta_{i,2} G_{i,2} - \beta_{ref,1}G_{ref,1} \\ +\beta_{ref,2} G_{ref,2} = f_{i,1} -f_{i,2} - f_{ref,1} +f_{ref,2}
		\end{split}		
		\end{equation}
		\begin{equation} \label{eq:finaldg}
		\begin{split}
			 \Delta G_{ij}(T) = k_B T \Big ( \Delta f_{ij}(T) - \Delta f_{ij}(T_{ref}) \Big ) + \\
			\frac{T}{T_{ref}} \Delta G_{ij}(T_{ref}) 	
		\end{split}
    		\end{equation} 
 Once the Gibbs free energy differences between polymorphs have been calculated, a set of coexistence pressures and temperatures are then determined from these free energy differences. The coexistence lines in the phase diagram are the intersections of the surfaces formed by the set of free energy differences. The points used to determine these lines are found by interpolation. First, the lowest Gibbs free energy, and thus the most stable, polymorph is determined at each $(T,P)$ point. Next, each combination of adjacent $(T,P)$ points is checked. If the most stable polymorph is not the same at any set of adjacent $(T,P)$ points, then a coexistence point must lie between them. Interpolation is then used with Eqs.~\ref{eq:coex} and~\ref{eq:coexP} to find where between the two points the coexistence point should lie. To make interpolation easier, the initial set of temperatures and pressures that are simulated are placed in a grid, although this is not strictly required. This approach can be seen schematically in Fig~\ref{fig:interp}.    

	\begin{equation} \label{eq:coex}
		T^{coex} = T_{1} + \frac{(T_{2}-T_{1})(\Delta G_{1,1}- \Delta G_{1,2})}{\Delta G_{1,1} - \Delta G_{2,2} - \Delta G_{1,2} + \Delta G_{2,2}}
	\end{equation}
	\begin{equation} \label{eq:coexP}
		P^{coex} = P_{1} + \frac{(P_{2}-P_{1})(\Delta G_{1,1}- \Delta G_{1,2})}{\Delta G_{1,1} - \Delta G_{2,2} - \Delta G_{1,2} + \Delta G_{2,2}}
	\end{equation}
The initial set of grid points can be chosen in a variety of ways. For the purposes of testing in this paper, no previous knowledge about the region of coexistence was assumed. Thus the initial points were set in a grid covering the entire region of interest for the phase diagram. However, if some previous knowledge of coexistence is available, gridpoints can be chosen around the roughly known phase equilibration line, or the phase diagram can be built out from the known region to encompass the region of interest. The only strict requirement is that there must be phase space overlap between regions of sampled states. Using approximately known coexistence region as a starting point increases the efficiency of the SIMR method by eliminating the need for simulations in regions of the phase diagram far from the coexistence lines. 

\begin{figure}
\begin{center}
\centering
\subfloat[]{\centerline{\includegraphics[width=0.5\textwidth]{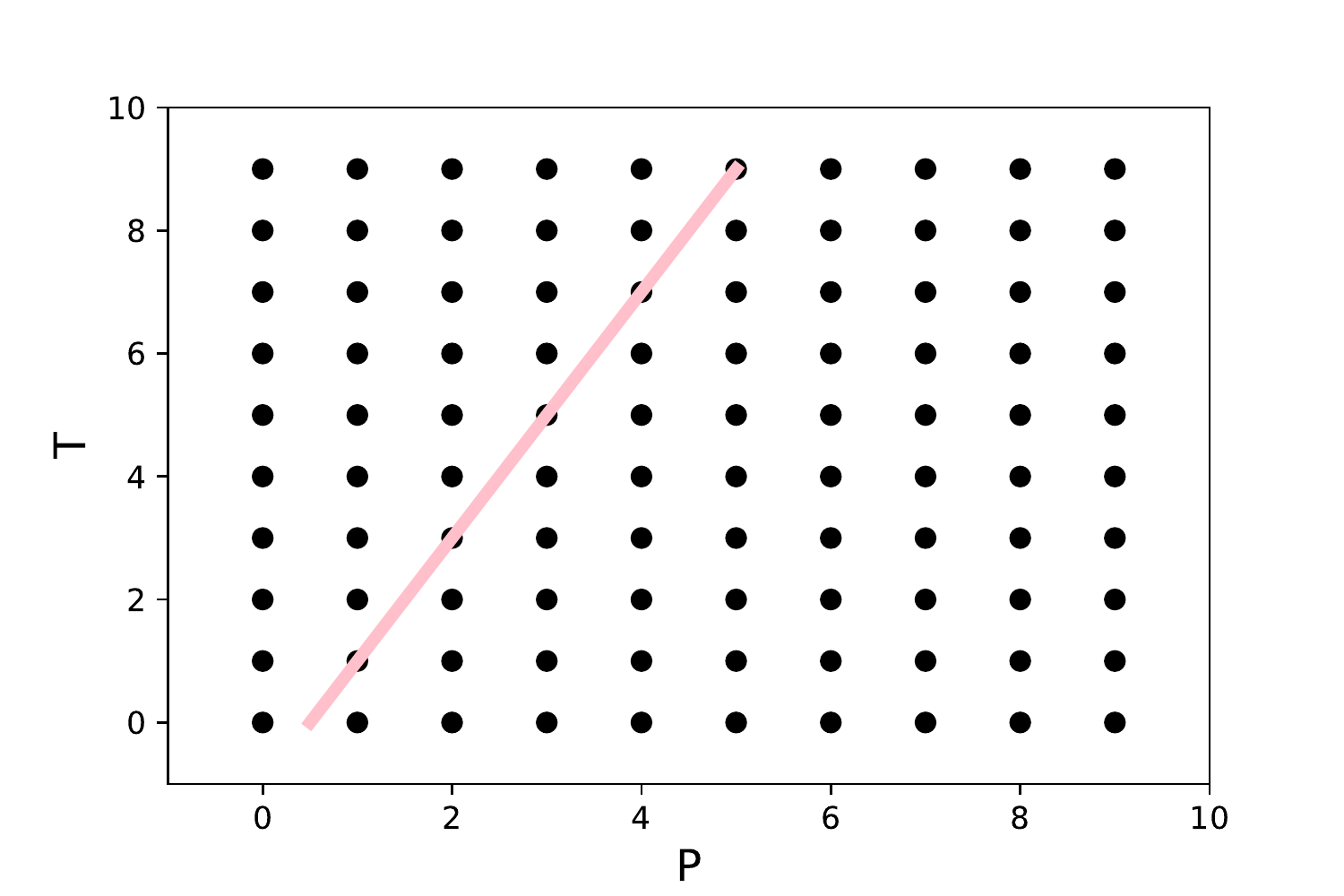}}}
\newline
\subfloat[]{\centerline{\includegraphics[width=0.5\textwidth]{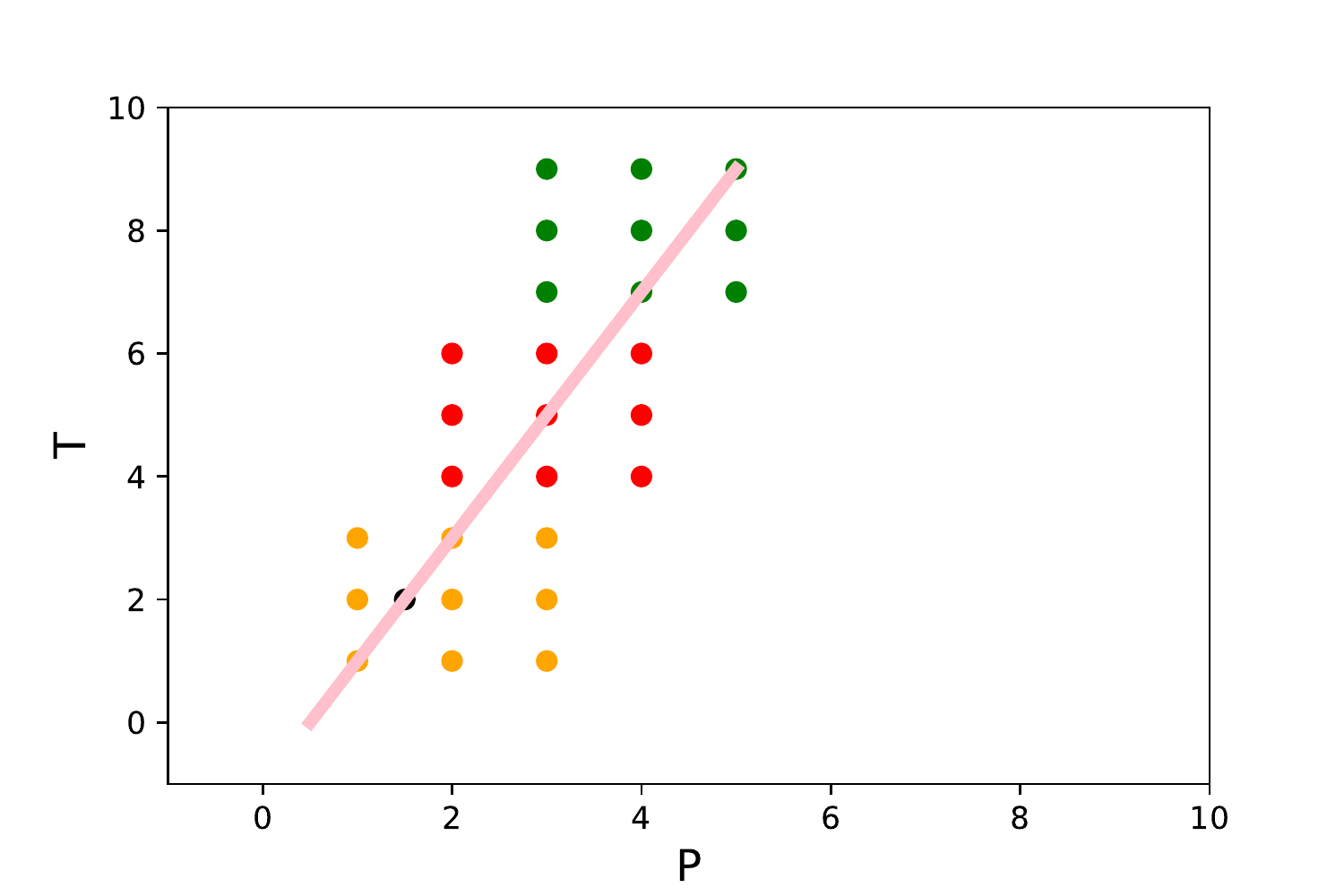}}}
\newline
\subfloat[]{\centerline{\includegraphics[width=0.5\textwidth]{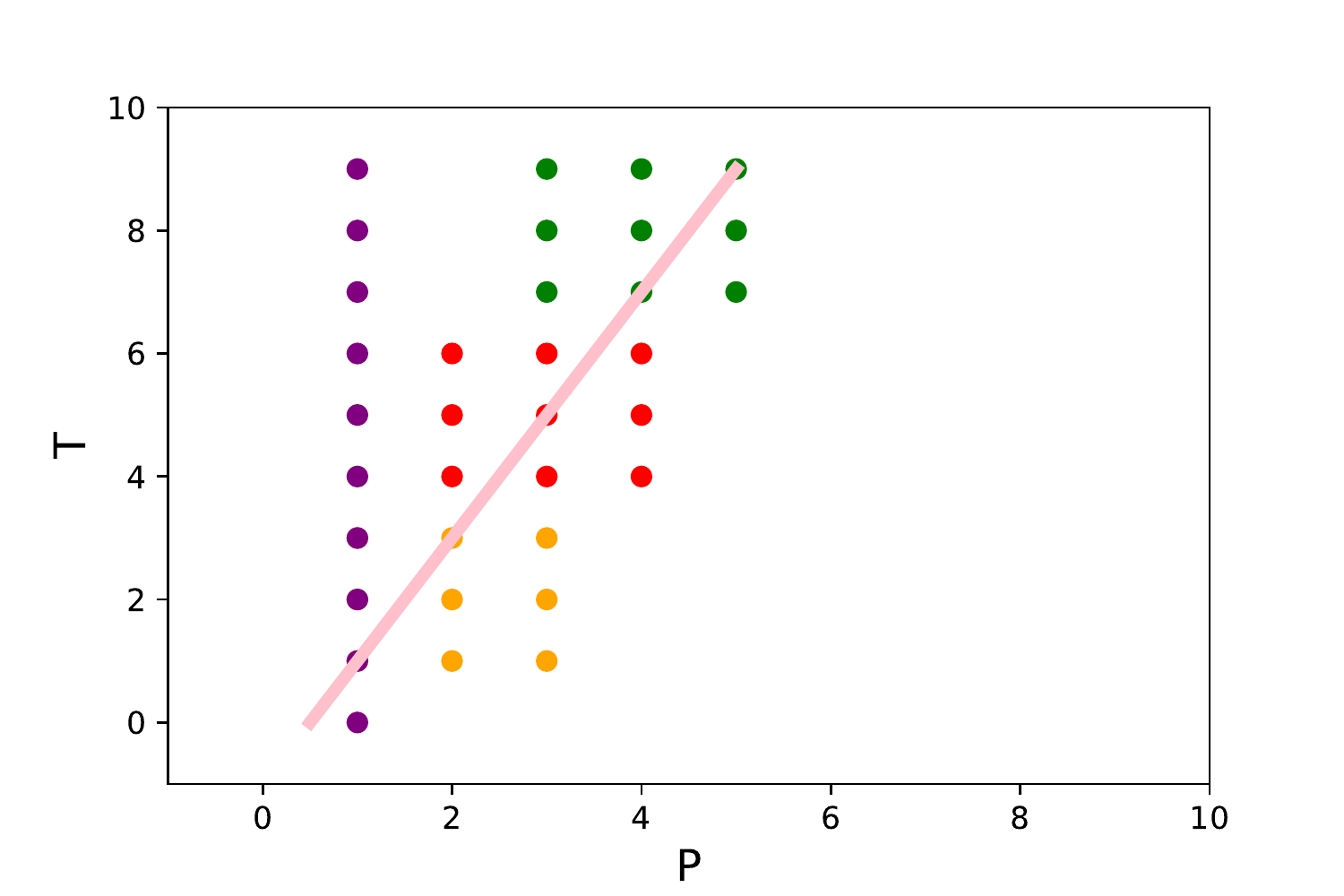}}}
\newline
        \caption{\label{fig:states} There are multiple ways of determining which $(T,P)$ states to simulation to generate the phase diagram. In (a) an evenly spaced grid is used. In (b) the initial point (black) was known and the simulations were placed in the region surrounding in the order of blue, red, green, to build the true line (yellow). In (c) the initial point was found by scanning a single pressure (purple) and then building out from the first determined coexistence point as in (b).}
\end{center}
\end{figure}

Three different ways of choosing the initial states are shown in Fig.~\ref{fig:states}. In (a), the simulated states were chosen to be evenly spaced in a grid over the entire region to be studied. This represents the case with no previous knowledge of any coexistence. Case (b) represents the case where a single coexistence point was known. In this case, sampled states were added around the initial point to determine the direction of the line at that point. After the first set of states are added, more can be added in the direction where the coexistence line that has been determined up to that point. This can be repeated until the entire line is determined. The third case, (c), is an example where a coexistence pressure is known but not the corresponding temperature or vice versa. In this case, the corresponding temperature can be found by simulating at multiple temperatures and the coexistence pressure. Multistate reweighting is then used to find which temperature corresponds to the coexistence pressure. Once this is done, the line can be built the same way as in case (b). 

One advantage of the SIMR method is that the error in the phase coexistence line can be estimated directly from the error in the reduced free energy difference, estimated by the MBAR approach (and implemented in the \texttt{pymbar} package). The uncertainty in the phase boundary line is a function of the value of the free energy difference and the slope of the free energy difference surface. The uncertainty in each of the reduced free energy difference values is computed by \texttt{pymbar}.  First, a simple error propagation is performed on the definition of Gibbs free energy difference found in Eq.~\ref{eq:finaldg}, where the uncertainty in the reduced free energy differences and the reference Gibbs free energy difference is used. This results in Eq.~\ref{eq:ddg} where  $\delta f_{1,ref} $ is the uncertainty in the reduced free energy of state 1 at the reference point.
\begin{equation} \label{eq:ddg}
\begin{split}
\delta G = [(\delta G_{ref} \frac{T}{T_{ref}})^{2} + (k_{B} T \delta f_{1,ref})^{2} +  (k_{B} T \delta f_{1,i})^{2} + \\
 (k_{B} T \delta f_{2,ref})^{2} + (k_{B} T \delta f_{2,i})^{2} ]^{1/2}
\end{split}
\end{equation}

\begin{figure}
\begin{center}
 \includegraphics[width=0.5\textwidth]{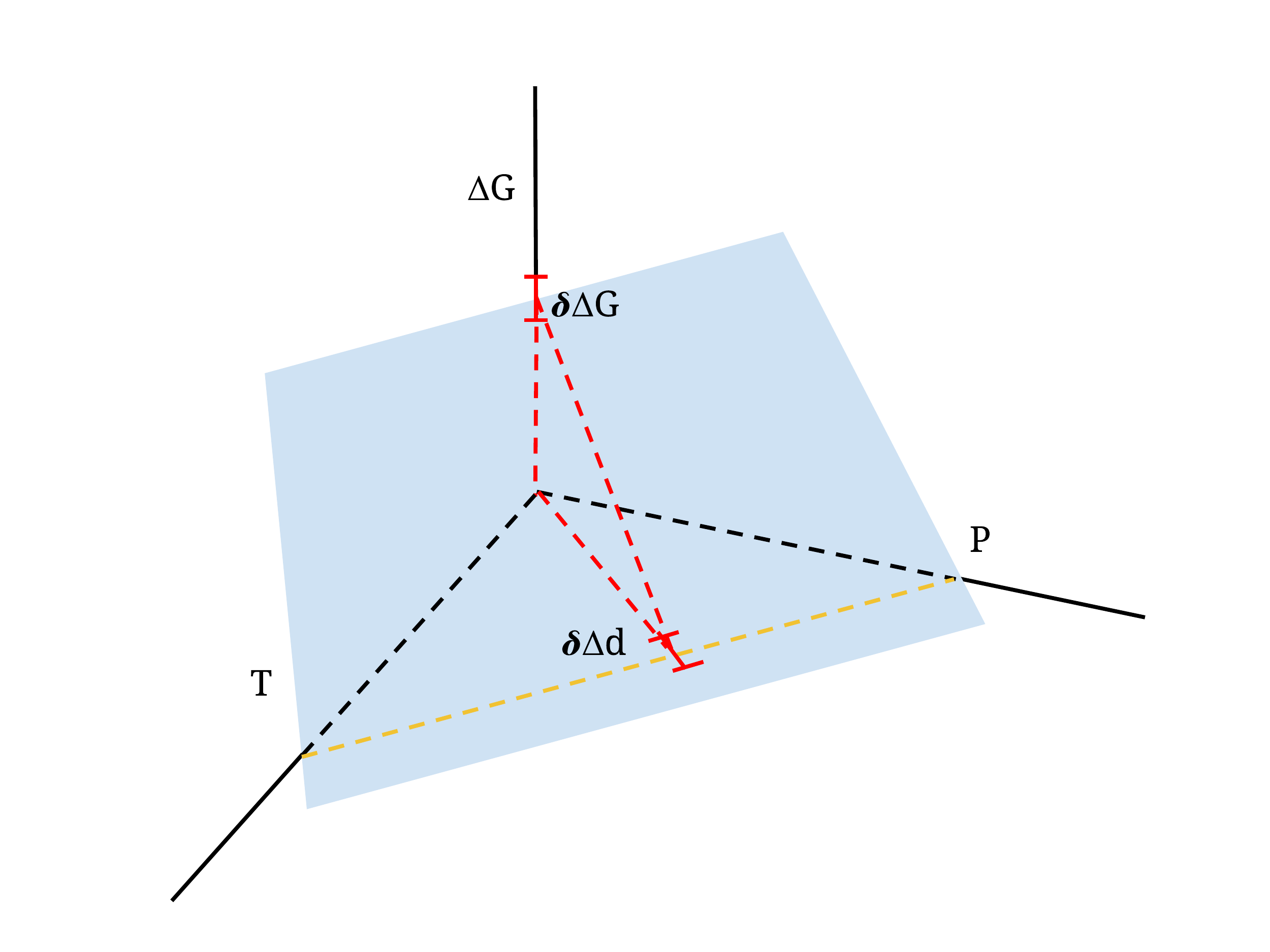} 
        \caption{\label{fig:errorfig} The uncertainty in the coexistence line perpendicular to the slope of the line is dependent on the uncertainty in the Gibbs free energy difference and the slope of the free energy surface. }
\end{center}
\end{figure}

This Gibbs free energy difference uncertainty value can then be used to calculate the uncertainty in the value of the coexistence line. To do this, the slope of the free energy difference surface must be calculated as a function of $T$ and $P$. The magnitude of the uncertainty in the coexistence line, $\delta d$, is perpendicular to the slope of the coexistence line at that point, and is given by Eq.~\ref{eq:uncertainty}. This can be seen in Fig.~\ref{fig:errorfig}:
\begin{equation} \label{eq:uncertainty}
\delta d = \sqrt{\left(\frac{\partial \Delta G}{\partial P}\right)^{2}+\left(\frac{\partial \Delta G}{\partial T}\right)^{2}} \delta \Delta G
\end{equation}
Once the set of coexistence points and their associated uncertainty have been determined, additional simulations in each polymorph can be performed at the $(T,P)$ values of each of the predicted coexistence points. The SIMR process will produce a set of $(T,P)$ coordinates. Additional simulations can be performed at these $(T,P)$ points. These simulations can then be incorporated into the multistate reweighting calculation as sampled states. These additional states serve two purposes. First, additional information in the region of coexistence will decrease the analytical uncertainty. Second, the spacing between the sets of $(T,P)$ points near the coexistence line will be smaller. This makes the interpolation used to find coexistence points more accurate. This process of adding new sampled states and recalculating the coexistence line can be repeated until a desired uncertainty in the line is reached. 

Due to the requirement that adjacent simulations have sufficient phase space overlap, the number of simulations performed is dependent on the width of the potential distributions of the simulations. Systems with wider potential energy and volume distributions can have larger spacing and still achieve phase space overlap. The width of these distributions, and thus the spacing in temperature and pressure that is allowable between simulations, depends on factors such as the temperature, pressure, and the size and flexibility of the molecule.

\subsection{Simulation Details}

\begin{figure}
\begin{center}
	\subfloat[I]{\includegraphics[width=0.15\textwidth]{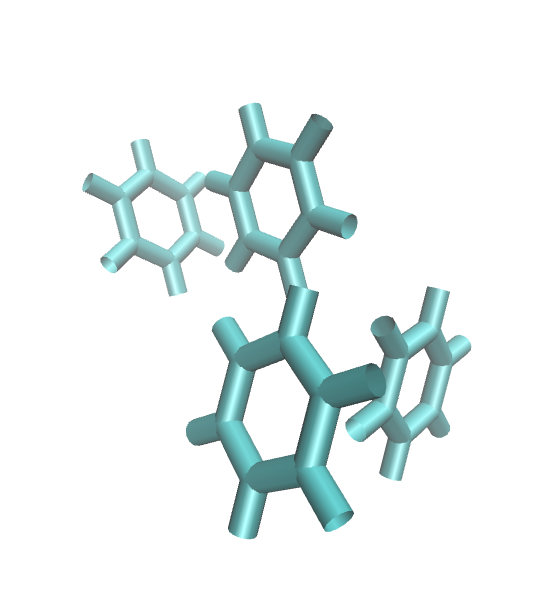}}
	\subfloat[II]{\includegraphics[width=0.15\textwidth]{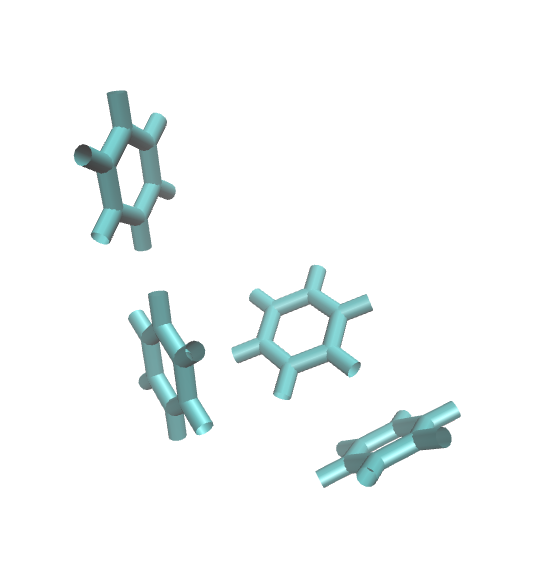}}
	\subfloat[III]{\includegraphics[width=0.15\textwidth]{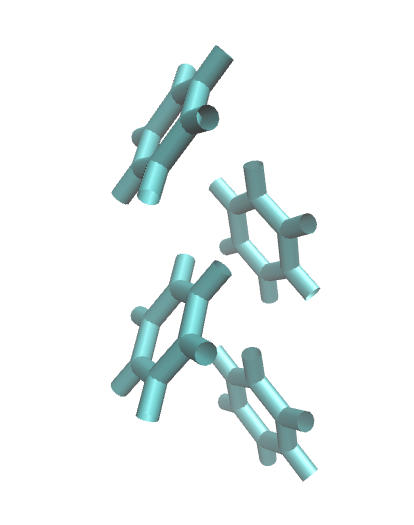}}
        \caption{\label{fig:bnzstruct} The three different polymorphs of benzene used in this study are I, II, and III.}
\end{center}
\end{figure}

To implement the SIMR method, 
we chose benzene as a test system.
Benzene 
is a small, rigid, well-studied organic molecule, and has at least three polymorphs which have been studied and observed experimentally~\cite{Raiteri2005b,  Thiery1988, Cansell1993a}. All simulations were performed using GROMACS 5.0.4~\cite{Berendsen1995} on the Bridges computational cluster \cite{Nystrom2015, Towns2015}. Each benzene simulation was run using a system of 4 independent benzene molecules. Since GROMACS has the requirement that the cell size be larger 1.5 times the cutoff distance, a supercell of 72 benzenes was simulated. 
A modification to GROMACS was used to average the forces on each unit cell within the supercell so that each individual unit cell moves identically. This modification is available as a branch from the main GROMACS git repository~\cite{gromacs_forceaverage}. This reduced the number of independently moving benzenes from 72 to 4, essentially simulating a single unit cell. We studied three polymorphs, benzene I, II, and III, used. The three polymorphic structures of benzene can be seen in 
Figure~\ref{fig:bnzstruct}.  
Simulations for the benzene phase diagram were performed every 700 bar between 1 and 55000 bar. The upper value of this range was chosen to be 10000 bar above the experimentally determined coexistence between polymorph I and II based on Raiteri et al.~\cite{Raiteri2005b} The temperature range for the simulations was between 60 and 280K at a spacing of 40K. This was chosen to avoid the melting point of benzene at 1 bar, which is approximately 278K~\cite{Menzies1932}. Spacing in the temperature and pressure directions were determined using the energy and volume distributions at their narrowest points. 

In all benzene simulations, the OPLS-AA potential was used~\cite{Damn1997}. This potential was previously shown to produce the correct polymorph stability ordering at 200K and 1 bar~\cite{Dybeck2016}. First, the system was equilibrated for 0.5 ns using anisotropic Berendsen pressure coupling~\cite{Berendsen1984} and a 1000 ps time constant. This allowed the simulation to equilibrate using a relatively stable pressure coupling. Following equilibration, production simulations were run for 4 ns each. 
The Parrinello-Rahman barostat was used for production~\cite{Parrinello1980}, which gives the proper fluctuations in volume for the NPT thermodynamic ensemble.

For all benzene simulations, Langevin dynamics was used for integration of the molecular dynamics simulations~\cite{langevin}.  Long range electrostatic interactions were handled using Particle Mesh Ewald~\cite{Essman1995} switch and a cutoff distance of 0.7 nm. Van der Waals interactions were treated with the PME Potential-Shift method with a cutoff of 0.7 nm. A Fourier spacing of 0.13 nm was used. 
A previous study showed that 0.7 nm cutoffs that included PME treatment of Lennard-Jones interactions were sufficient for quantitative calculations of benzene polymorph stability~\cite{Dybeck2016}.

\section{Results}

\subsection{Full Molecular Dynamics Phase Diagram of Benzene}

\begin{figure*}
\begin{center}
\subfloat[]{\includegraphics[width=0.50\textwidth]{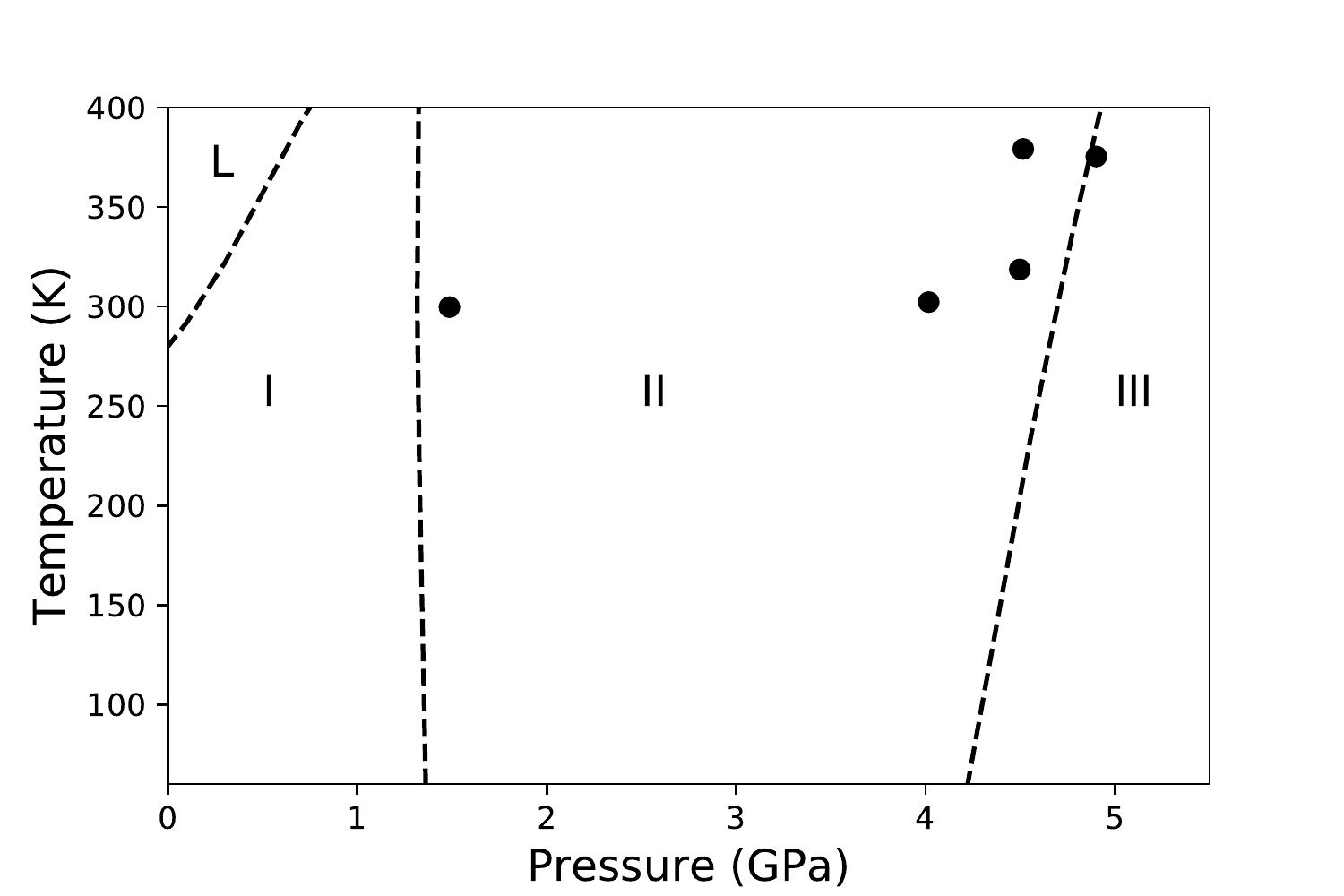}}
\subfloat[]{\includegraphics[width=0.48\textwidth]{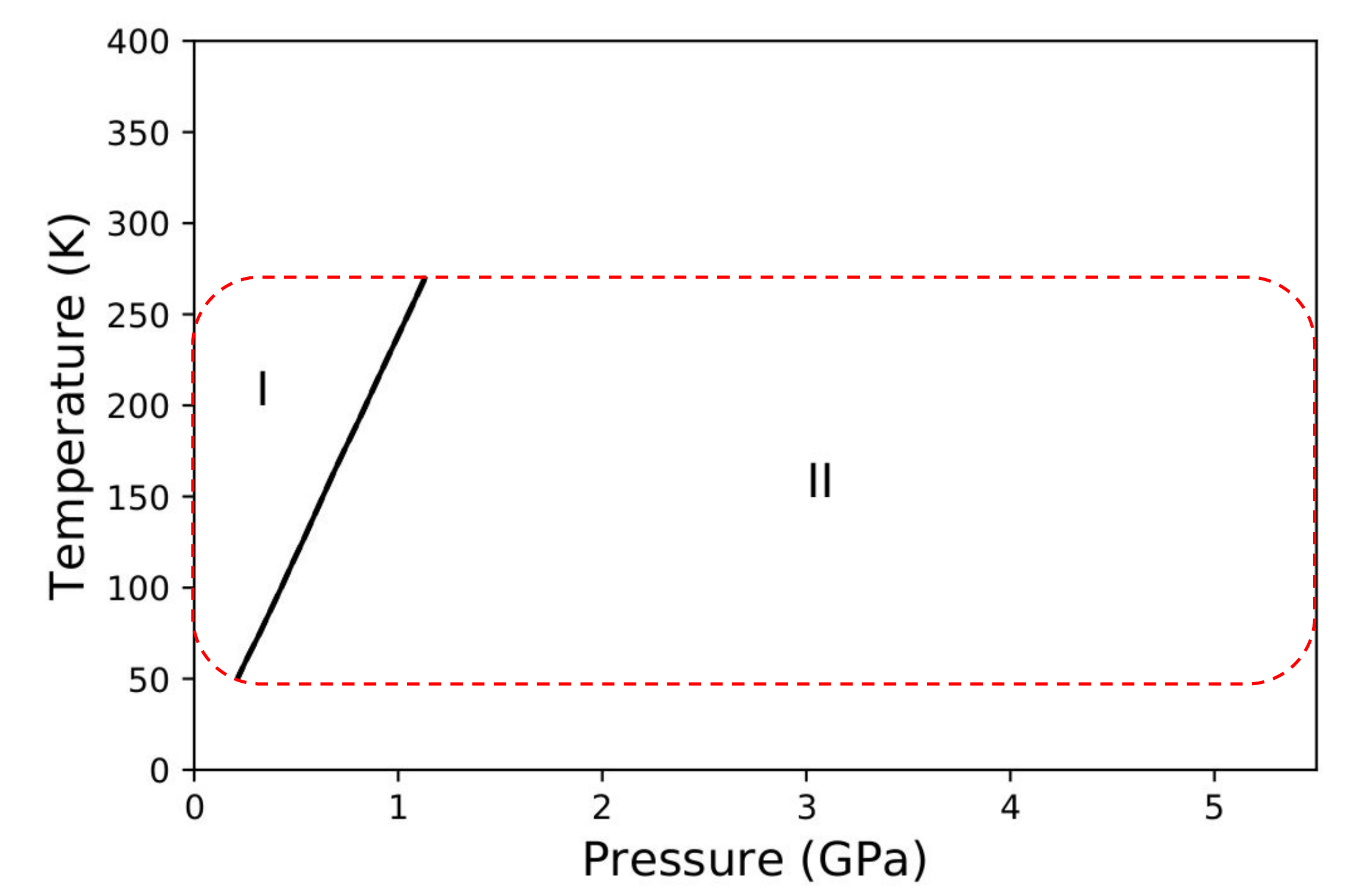}}
	\caption{\label{fig:bnzcomp} (a) Coexistence points and the assumed coexistence lines (dotted) of benzene generated using experiment and (b) the region of simulation (red dotted line) and coexistence lines obtained with SIMR show agreement in the ordering of benzene I and II but not quantitative agreement. Experimental results figure and data adapted from Raiteri et al.~\protect{\cite{Raiteri2005b}}}
\end{center}
\end{figure*}

\begin{figure}
\centering
	\includegraphics[width=0.55\textwidth]{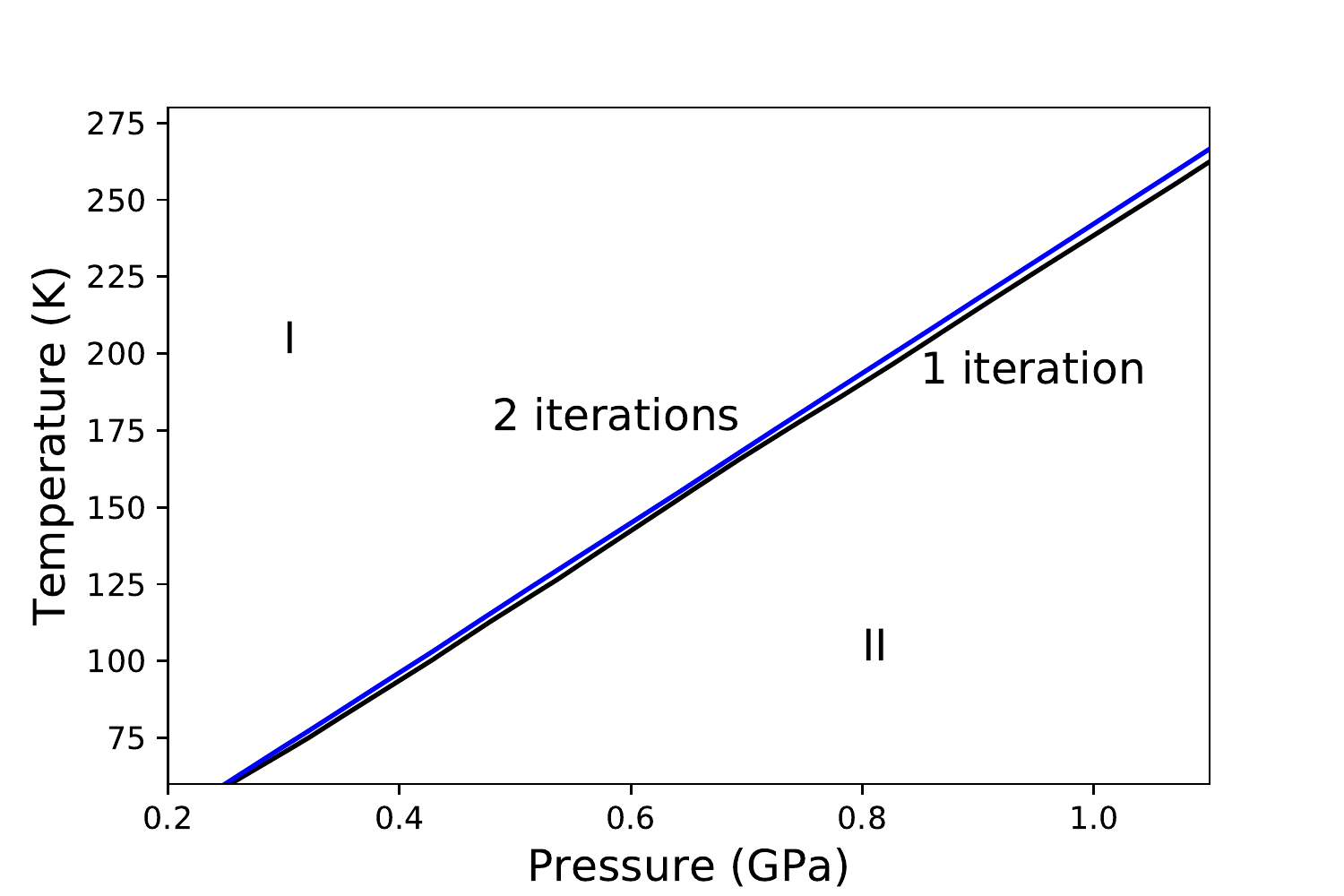}
	\caption{\label{fig:benzene_added} The difference between the predicted coexistence line with SIMR using one and two iterations of sampled states shows minor differences.}
\end{figure}

Using the SIMR method, we present the first computationally predicted solid phase diagram of crystalline benzene in Fig.~\ref{fig:bnzcomp}. This phase diagram studies benzene in the entire region between 0.0001 to 5.5 GPa and 60 to 280 K. This phase diagram shows strong pressure dependence and weak temperature dependence. In comparison to experimental results for the phase diagram of benzene, the ordering of polymorphs and transition between phase I and II is qualitatively the same \cite{Raiteri2005b, Ciabini2005}. Quantitatively, the transition between I and II occurs at a higher pressure experimentally than in the phase diagram predicted using SIMR. A comparison between the previous experimental results and the SIMR results is shown in Fig.~\ref{fig:bnzcomp}. In previous experimental work, the lowest experimentally determined point is 300K, the coexistence line below that point is an extrapolation. This may account for some of the differences between SIMR and experiment. The highest value chosen for this phase diagram was chosen to be 280K, in order to avoid potential melting during the simulations. In order to refine the line and determine the magnitude of the effect of adding extra iterations of the SIMR method, two iterations of sampled states were used. The difference for a portion of the coexistence line when adding extra sampled states based on the initial coexistence line can be seen in Fig.~\ref{fig:benzene_added}. The ordering of polymorphs as a function of pressure is consistent with the results of Schnieder et al.~\cite{Schnieder2016}

\subsection{Error Analysis}

One advantage to the SIMR method is that the calculation of the uncertainty in the coexistence line is straightforward and computationally cheap. One of the outputs of the \texttt{pymbar} package is a matrix consisting of the uncertainty in the free energy between each combination of states, as calculated by the covariance matrix in the MBAR calculation. This uncertainty, can be propagated through the Gibbs free energy difference Eq.~\ref{eq:finaldg} to produce Eq.~\ref{eq:ddg}. The resulting uncertainty is the uncertainty in the free energy difference between polymorphs. However, the desired uncertainty is in the position of the coexistence line. This uncertainty in coexistence perpendicular to the line is determined using Eq.~\ref{eq:uncertainty}. A subsections of the benzene phase diagram where the uncertainty lines can be discerned is shown in Figure 15. Statistical bootstrapping, with 100 bootstrap samples, was performed on the configuration input to \texttt{pymbar} and the uncertainty determined by bootstrapping agreed to within twenty percent of the analytical uncertainty. This indicates that the faster analytical error determination is sufficiently accurate and should be used. 
Since each bootstrap sample requires recalculation of the reduced free energies and full solution of the nonlinear MBAR equations, it is computationally favorable to use analytically obtained uncertainties.
\begin{figure}
\begin{center}
\centering
	\subfloat[]{\centerline{\includegraphics[width = 0.5\textwidth]{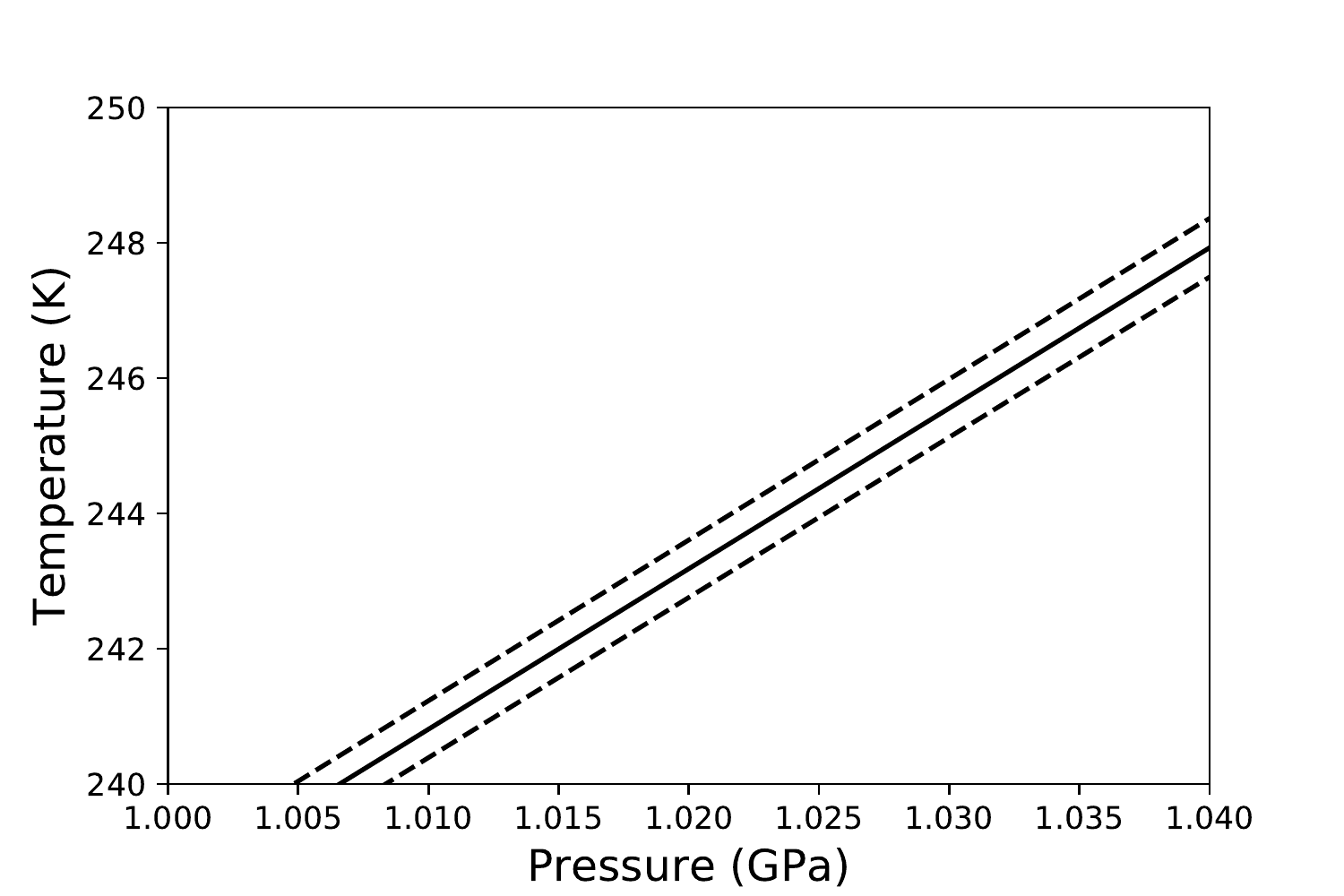}}}
	\caption{\label{fig:bnzerr} A subsection of the benzene phase diagram allows the uncertainty in the coexistence line to be visualized with dashed lines.}
\end{center}
\end{figure}

\subsection{Dependence of Efficiency on System Size}

An important problem that has been brought up with reweighting
approaches is the poor scaling scaling of the method with increasing system size \cite{Bruckner2010}.  As the
size of systems increases, the energy distributions narrow. This
means that reweighting becomes rapidly less efficient as overlap
decreases, in most cases exponentially quickly with size.

It is therefore important to examine how SIMR scales with system size.  As seen
in Eq.~\ref{eq:uncertainty}, the statistical error in the phase diagram line
is directly proportional to magnitude of error in $\delta \Delta
G_{ij}$. We first make the approximation that when finding the value of an intersection point, only two states are primarily
responsible; the two states that are being interpolated between to find the intersection point.  We can then use a simplified two state
system that is easier to analyze quantitatively.

With equal numbers of samples, $N_{samples}$, from each state, then the
uncertainty is equal to Eq.~\ref{eq:vartooverlap}, where $O$ is the overlap
integral, Eq. \ref{eq:size2} as derived by Bennett~\cite{Bennett1976}.
\begin{equation}
\var_{\Delta f}(N_{samples}) = \frac{\left(O_{ij}^{-1}-2\right)}{N_{samples}} \label{eq:vartooverlap}
\end{equation}

\begin{equation} \label{eq:size2}
O_{ij} = \int \frac{P_i(\vec{x}) P_j(\vec{x})}{P_i(\vec{x}) + P_j(\vec{x})}d\vec{x}.
\end{equation}

Assuming that the distributions are two harmonic oscillators
with the same force constant $k$, and the means are separated by $c$, we can
then plug the distributions
$P_{i}(x)=\sqrt{\frac{k}{2\pi}}e^{-\frac{k}{2}(x-c/2)^2}$ and
$P_{j}(x)=\sqrt{\frac{k}{2\pi}}e^{-\frac{k}{2}(x+c/2)^2}$, into
Eq.~\ref{eq:size2}, and simplify this integral to Eq.~\ref{eq:simple}.

\begin{eqnarray} \label{eq:simple}
O_{ij}(c,k) = \sqrt{\frac{k}{8\pi}} e^{-\frac{kc^2}{8}} \int \frac{e^{-\frac{kx^2}{2}}}{\cosh(\frac{ckx}{2})} \vec{x}
\end{eqnarray}

However, this integral does not appear to have an analytical
solution.  We can rewrite the integral part of the above expression
as:
\begin{eqnarray*}
&=&\int \exp\left(-\frac{k}{2}x^2 -\ln\left(\cosh\left(\frac{ckx}{2}\right)\right)\right) dx
\end{eqnarray*}
And then rewrite in terms of a Taylor series:
\begin{eqnarray*}
&=&\int \exp\left(-\frac{k}{2}x^2 - \frac{k^2c^2x^2}{2} + \frac{k^4c^4x^4}{12} - \frac{k^6c^6x^6}{12} + \ldots   \right.
\end{eqnarray*}
We chose to Taylor expand the argument of the logarithm of the exponential
of the integrand instead of the integrand itself because we know that
a probability distribution must always be positive, which would not be true if we expanded the integrand itself.
Because the integral doesn't converge for the 2nd order term, and we
are only looking for leading term behavior, we truncate after the
first term in the Taylor series. This integral is now straightforward, and
yields the full overlap equation \ref{eq:fulloverlap}
\begin{eqnarray} \label{eq:fulloverlap}
O_{ij}(k,c) &=& \sqrt{\frac{k}{8\pi}} e^{-\frac{kc^2}{8}} \sqrt{\frac{8\pi}{k(4+c^2k)}} \\
                &=& \frac{e^{-\frac{kc^2}{8}}}{\sqrt{4+kc^2}}
\end{eqnarray}
It is important to note that this is only a function of $kc^2$, and not of
either of the variables individually. This makes sense in terms of
scaling, as $kc^2$ is unitless.  Thus, $kc^2$ can be replaced by a dimensionless parameter $a$ since the
overlap only varies with this combination of parameters $k$ and $c$.  If
we increase the number of harmonic oscillators further, then we know
the distribution will still be Gaussian (the sum of Gaussians is a
Gaussian). We will then replace $k$ with $k/N$, since the variance of
the distribution becomes larger by $N$, and $\sigma^2 = 1/k$. We also replace $c$
with $Nc$, since the means of the distributions are scaled by the
number of oscillators.  This means $kc^2=a$ is replaced with
$Nkc^2=Na$. We than then use equation ~\ref{eq:vartooverlap} to obtain equation \ref{eq:var1}.
\begin{equation} \label{eq:var1}
\var_{\Delta f}(k,c,N) \propto \frac{e^{Na/8}\sqrt{4+Na}-2}{N^2}
\end{equation}
The  $N^2$ factor in the denominator is because in finding the ``per
mol'' uncertainty, the standard deviation decreases by $N$, not
$\sqrt{N}$, as the value of the per mole uncertainty is completely
correlated with itself.

Finally, taking into account the value of
$N_{samples}$, the variance will be:
\begin{equation} \label{eq:var2}
\var_{\Delta f}(a,N,N_{samples}) = \frac{e^{Na/8}\sqrt{4+Na}-2}{N^2N_{samples}}
\end{equation}
The standard error in the estimate of the free energies is then equal to
\begin{equation}
\sigma_{\Delta f}(k,c,N,N_{samples}) = \frac{1}{N}\sqrt{\frac{e^{Na/8}\sqrt{4+Na}-2}{N_{samples}}} \label{eq:sigma_gauss}
\end{equation}

We can now qualitatively answer the question of how the efficiency of
the methods scales with $N$. Given a value of $a=kc^2$, the efficiency
can actually \emph{increase} as a function of $N$ (i.e. statistical
uncertainty decreases), until a minimum is reached,
at which point the statistical uncertainty increases rapidly.  We can
solve this numerically to find that the variance is minimized at
$Na=10.97$.  So given a value of $a$, we find that the simulation is
most efficient at $N = \frac{10.97}{a}$. To remain at this high efficiency point as $N$
increases, we need to decrease the spacing.  However, since $a=kc^2$, we find that we must have $c \propto N^{1/2}$. For harmonic
oscillators, at least, we find that we improve efficiency as $N$
increases to $N = 10.97/a$, and then we must adjust the spacing.

However, how does this finding translate into problems that are
\emph{not} 1-D harmonic oscillators?  For example, crystal systems are
usually composed of hundreds of atoms, so $\Omega(E)$ is much more
complex. However, it makes intuitive sense to treat a large collection
of systems as a Gaussian distribution, due to the law of large
numbers. Additionally, the positions of particles in a crystal can often be well approximated by a
harmonic distribution, so the underlying configurational distribution
is itself harmonic.

To determine how well this approximation translates, we attempt to fit realistic crystal simulations to the analytical
results obtained here. For this system, we don't have a good sense of
what either $k$, or especially, $c$ are. We can adjust the spacing not
in configurational direction, but rather in $T$ and $P$. We can,
however, gather data on the uncertainty as a function of $N$, and fit
to the non-dimensional parameter $a$. If the model is useful, we will
obtain good agreement between the data and the model.

For this test, all simulations are of the Lennard-Jones FCC phase, run in the LAMMPS package \cite{PLIMPTON1995}. The FCC structure itself was generated by the LAMMPS package and system sizes between 32 and 500 atoms were used. The cutoff was 2.5 $\sigma$ for all simulations and each simulation was run for 8 million reduced time steps.

\begin{figure*}
\begin{tabular}{cc}
\includegraphics[width=0.50\textwidth]{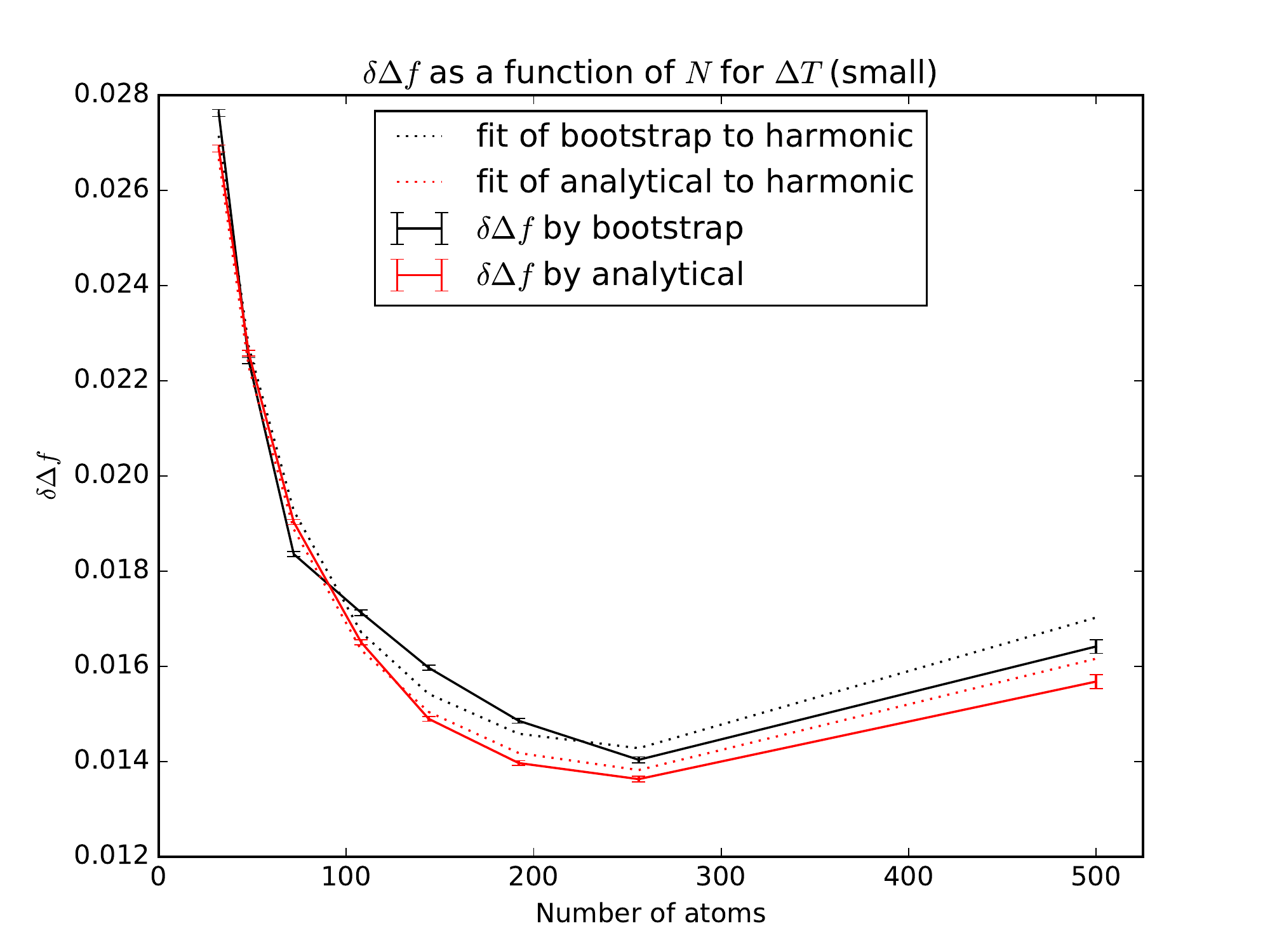} & \includegraphics[width=0.50\textwidth]{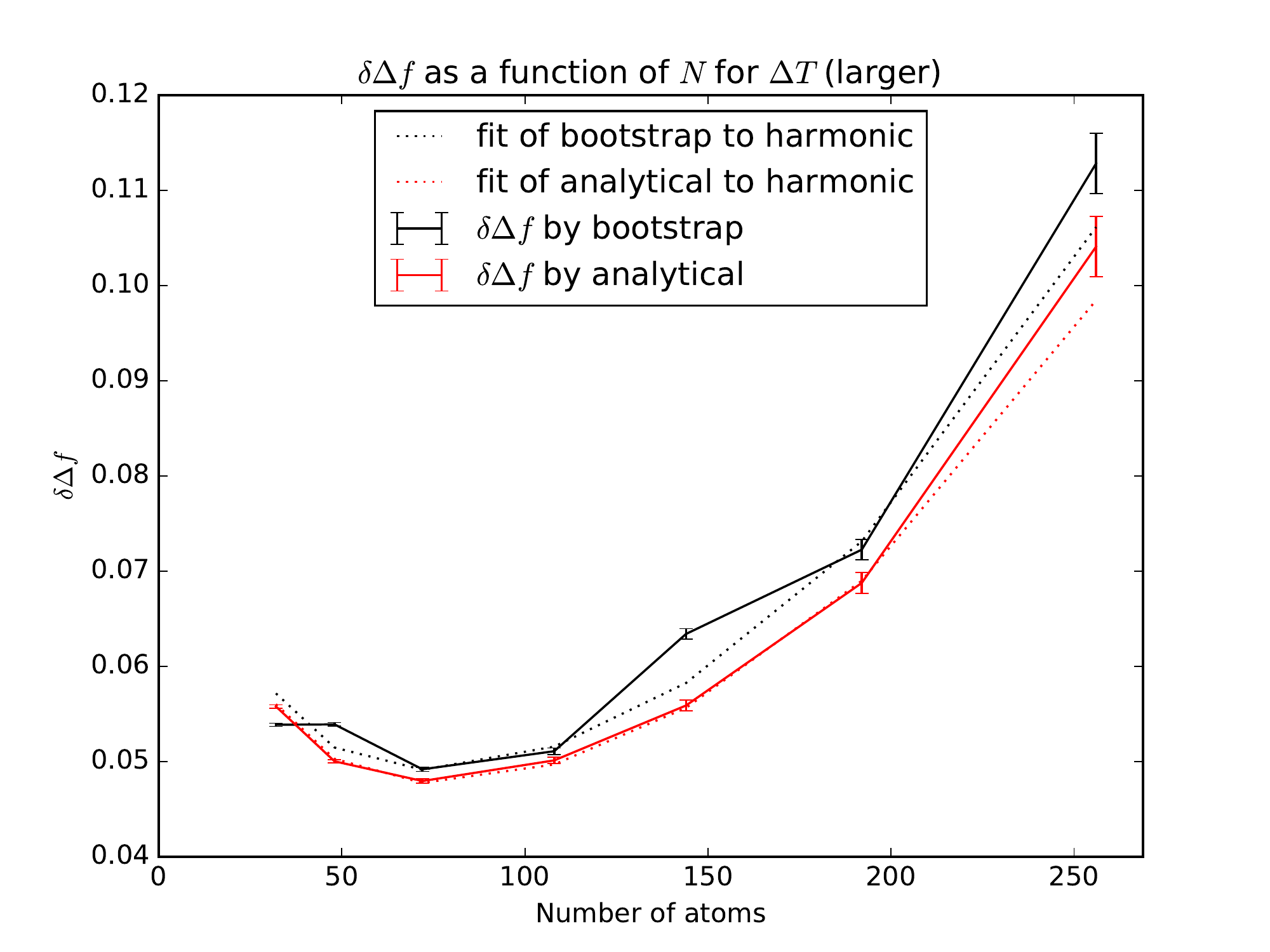}\\
(a) & (b) 
\end{tabular}
\caption{Uncertainty in the reduced free energy $f$ as a function of
  system size $N$ at $P^*=3.0$ between $T^*=0.30$ and $T^*=0.35$ in
  subfigure (a), and between $T^*=0.30$ and $T^*=0.40$ in subfigure
  (b). Uncertainty is estimated in two ways: (1) (green line) using
  the analytical error estimate for BAR (the two-state version of
  MBAR) and (2) (black line) using the bootstrap estimate of the free
  energy with 500 bootstraps.  We also show the fit to Eq.~\protect{\ref{eq:sigma_gauss}}, 
  the harmonic approximation.\label{fig:dTvsN}}
\end{figure*}

All analysis was done using the uncertainty in the reduced free
energy, which can then be propagated into the free energy difference
by Eq.~\ref{eq:ddg}. Fig.~\ref{fig:dTvsN}, shows the
uncertainty in the reduced free energy, $f$, at $P^*=3.0$ between
$T^*=0.30$ and $T^*=0.35$ in subfigure (a), and between $T^*=0.30$ and
$T^*=0.40$ in subfigure (b). Uncertainty is estimated in two ways: 1)
(green line) using the analytical error estimate for BAR (the
two-state version of MBAR)~\cite{Bennett1976} and 2) (black line)
using the bootstrap estimate of the free energy with 500 bootstraps.
We also show a fit to Eq.~\ref{eq:sigma_gauss}, with free
parameter $a$. The $N_{samples}$ is chosen as the mean of the number
of samples taken from each of the two states.

For differences in $T^*$, using the harmonic approximation to estimate
the statistical error as a function of $N$ works well, and both
bootstrap and analytical error estimates agree very well.  For the
$\Delta T^* = 0.05$ case (a), a nonlinear least squares fit (performed
with the \texttt{scipy} \texttt{optimize} module \texttt{curve\_fit}
function) gives $a = 0.0430 \pm 0.006$ for the bootstrap uncertainty
and $a=0.0417 \pm 0.003$ for the analytical error estimate, fitting
only to the $a$ parameter; visually, it is clear that the
uncertainties are in excellent agreement to the single harmonic
oscillator theory. 

For the $\Delta T^* = 0.10$ case, the nonlinear least squares fit
approach gives $a = 0.148 \pm 0.03$ for bootstrap and $a=0.1438 \pm
0.0004$ for analytical estimates, visually clearly a good
fit. Additionally, we find that $a(T^*=0.05)/a(T^*=0.10)$ is $3.45 \pm
0.14$, not entirely inconsistent with the idea that the harmonic
oscillator theory remains roughly true for far more complex solid
systems, where increasing the $\Delta T$ by 2 would increase $a$ by 4.

%Error propagation for ratios.
%$a = 0.1438/0.0417 = 3.45 \pm 0.14
%
% (0.003/0.0417)^2 + (0.0004/0.1438)^2
% (0.042)^2 + (0.0029)^2 = (0.042)^2
% da = 0.042 * 3.4 = 0.14

\begin{figure*}
\begin{tabular}{cc}
\includegraphics[width=0.50\textwidth]{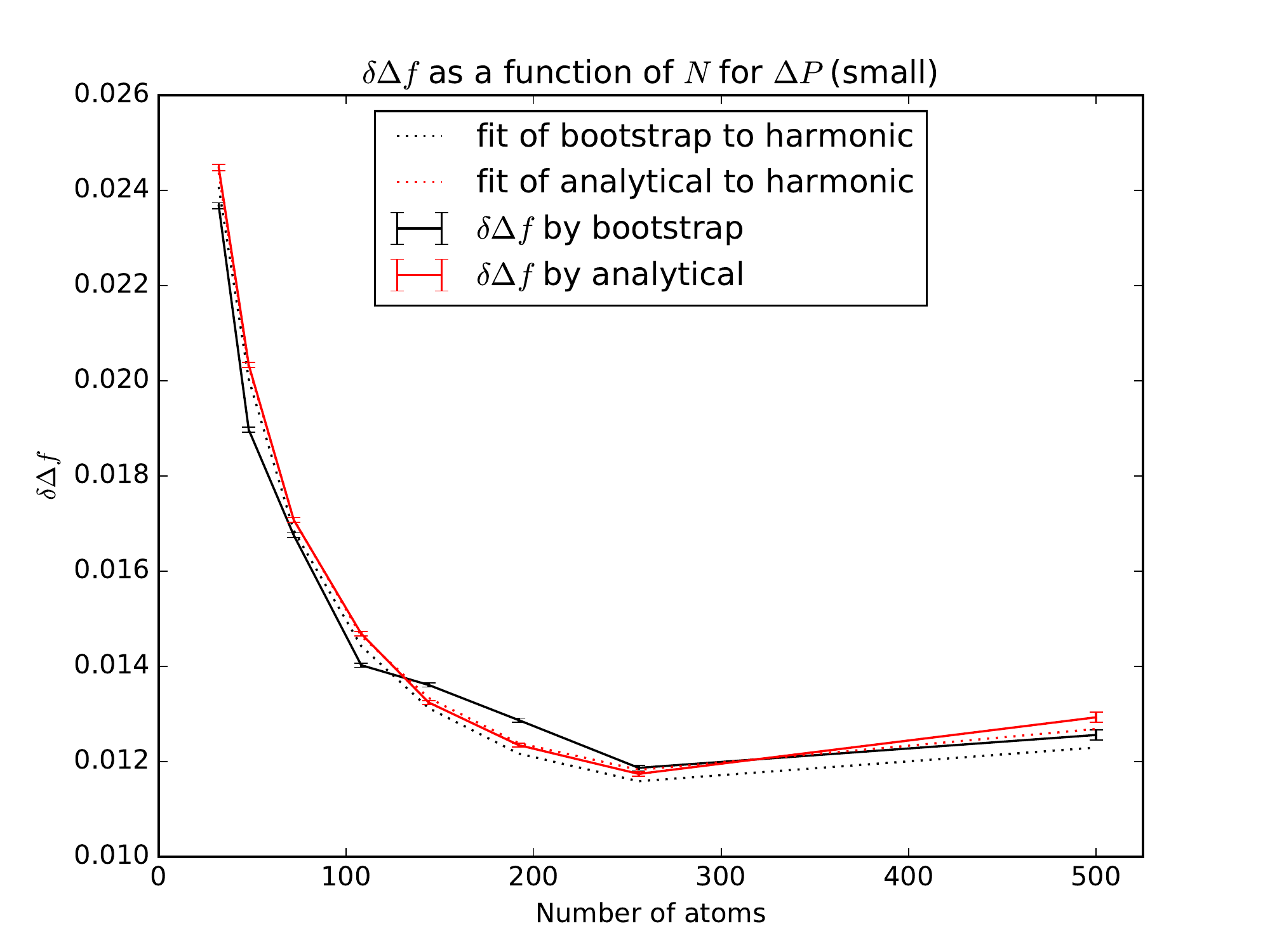} & \includegraphics[width=0.50\textwidth]{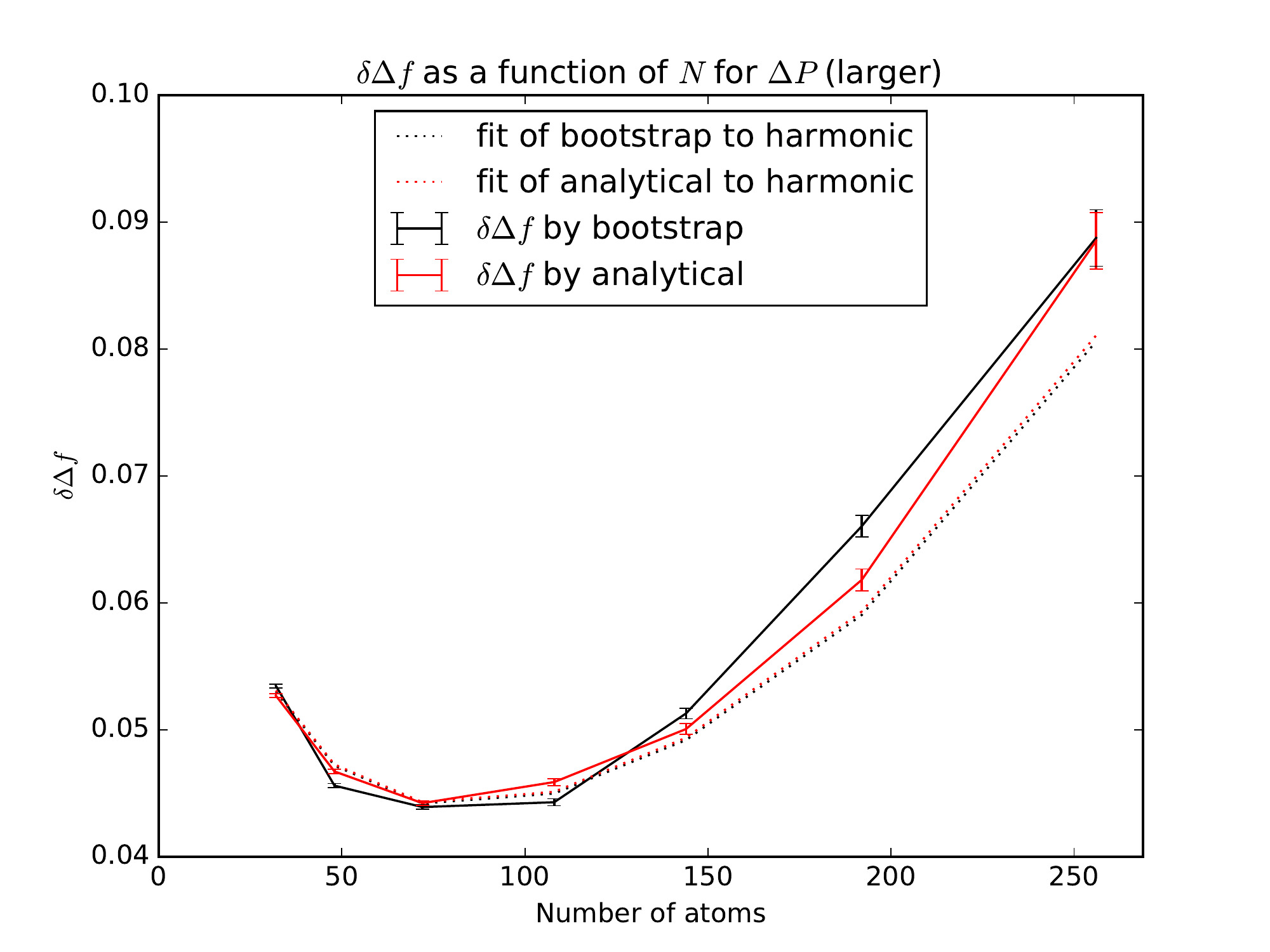}\\
(a) & (b) 
\end{tabular}
\caption{Uncertainty in the reduced free
energy $f$ as a function of system size $N$ at $T^*=0.35$ between
$P^*=2.0$ and $P^*=3.0$ in subfigure (a), and between $P^*=2.0$ and
$P^*=4.0$ in subfigure (b). Uncertainty is estimated in two ways: (1)
(green line) using the analytical error estimate for BAR (the
two-state version of MBAR) and (2) (black line) using the bootstrap
estimate of the free energy with 500 bootstrap samples.  We also show the fit
to the harmonic result in Eq.~\protect{\ref{eq:sigma_gauss}}.\label{fig:dPvsN}}
\end{figure*}

For differences in $P^*$, the match to harmonic theory is even more
accurate. Figure~\ref{fig:dTvsN}, shows the uncertainty in the reduced
free energy $f$ at $T^*=0.35$ between $P^*=2.0$ and $P^*=3.0$ in
subfigure (a), and between $P^*=2.0$ and $P^*=4.0$ in subfigure
(b). Uncertainty is again estimated in two ways: 1) (green line) using
the analytical error estimate for BAR (the two-state version of
MBAR)~\cite{Bennett1976} and 2) (black line) using the bootstrap
estimate of the free energy with 500 bootstrap trials.  Figure
\ref{fig:dTvsN} also shows a fit to equation ~\ref{eq:sigma_gauss},
where again the two free parameters is only $a$.  The number of
samples is estimated as the mean of samples from both sampled states.

For differences in $P^*$, using the harmonic approximation to estimate
the statistical error as a function of $N$ works well, and both
bootstrap and analytical error estimates agree very well.  For the
$\Delta P^* = 1.0$ case (a), a nonlinear least squares fit (performed
with the \texttt{scipy} \texttt{optimize} module \texttt{curve\_fit}
function) gives $a = 0.0345 \pm 0.001$ for the bootstrap uncertainty
and $a=0.0353 \pm 0.0001$ for the analytical error estimate, fitting
only to the $a$ parameter; Visually, the fit is excellent.

For the $\Delta P^* = 2.0$ case, the nonlinear least squares fit
approach gives $a = 0.133 \pm 0.002$ for bootstrap and $a=0.133 \pm
0.001$ for analytical estimates. Additionally, we find that
$a(P^*=1.0)/a(P^*=2.0)$ is $3.8 \pm 0.1$, indicating even more clearly
that the findings for harmonic oscillators remain roughly true for far
more complex solid systems under pressure changes.  

In all cases, we see that the anaytically estimated uncertainty is
very closely approximated by the significantly more expensive
bootstrap uncertainty.

%Error propagation for ratios.
% a = 0.133/0.035 = 3.8 \pm 0.1
%
% (0.001/0.0345)^2 + (0.002/0.133)
% (0.029)^2 + (0.015)^2 = (0.033)^2
% da = 0.033 * 3.8 = 0.1

\begin{figure} 
\centering
	\includegraphics[width=0.5\textwidth]{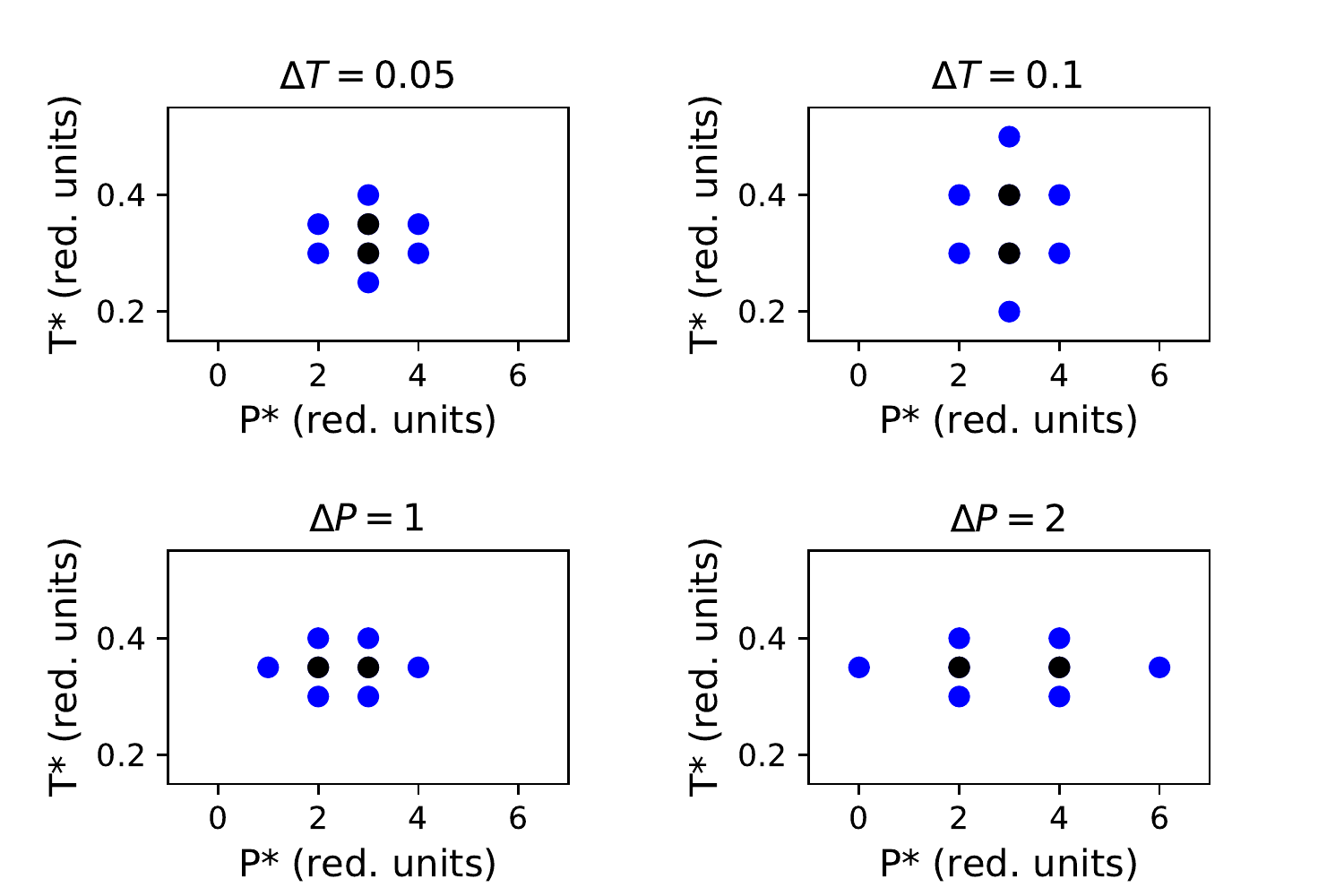}
	\caption{\label{fig:placing} For each comparison of size dependence, the free energy between two adjacent states (black) was studied, and information from adjacent states (blue) was included.}
\end{figure}

For the reweighting approaches described in this paper, we generally
use MBAR, which predicts free energies at all available collected
states.  For each of the cases above, we add six states, the nearest
neighbors in the grid space. The placement of the simulated states for
all cases can be seen in Table~\ref{table:spacings} and illustrated in
Fig. ~\ref{fig:placing}.  We find that including the additional
`diagonal' states, which differ from the two `central' states in both
$T$ and $P$ for a total of 12 states, changes the uncertainties and
free energies negligibly, and we thus analyze the size scaling of MBAR
with only the 6 additional nearest states, for at total of 8 states.

\begin{table*}
{\footnotesize
\begin{tabular}{|ccc|c|c|}
\hline
Quantity & $\Delta T^*$ grid & $\Delta P^*$ grid & 2 direct states & 6 nearest neighbor states   \\
$f(\Delta T^*)$ & 0.05 & 1 &   [0.30,3.0], [0.35,3.0]  & [0.30,2.0], [0.35,2.0], [0.30,4.0], [0.35, 4.0], [0.25,3.0], [0.40,3.0] \\
$f(\Delta T^*)$ & 0.10 & 1 &   [0.30,3.0], [0.40,3.0]  & [0.30,2.0], [0.30,2.0], [0.30,4.0], [0.40, 4.0], [0.20, 3.0], [0.50,3.0] \\
$f(\Delta P^*)$ & 0.05 & 1 &   [0.35,2.0], [0.35,3.0]  & [0.20, 2.0], [0.40, 2.0], [0.20, 3.0], [0.40, 3.0], [0.35, 1.0], [0.35, 3.0] \\
$f(\Delta P^*)$ & 0.05 & 2 &   [0.35,2.0], [0.35,4.0]  & [0.30,2.0], [0.40,2.0], [0.30,4.0], [0.40, 4.0], [0.35, 0.0], [0.25,6.0] \\ 
\hline
\end{tabular}
}
\caption{Choices of $T^*$ and $P^*$ for testing the size scaling of 2 state and 8 state reweighting.\label{table:spacings}}
\end{table*}

We note that for phase diagram determination, where we do not know
between which pairs of state points the phase intersection lies, MBAR
offers the additional advantage that it allows all states to be
determined simultaneously.  However, at this point, we are interested
in the estimates of the uncertainty, so we can take a minimal number
of samples that appear to contribute to a significant extent.

One challenge in fitting equation \ref{eq:sigma_gauss} is that with MBAR, it
is no longer quite clear what should be used for $N_{samples}$: all of
the samples at all of the states in MBAR, even when many of them are
not directly interacting? We choose instead to use the mean of the
number of samples from the two states also used in BAR.  This has the
advantage that the standard errors are directly comparable; the ratio of the uncertainties between
the uncertainty in MBAR and BAR is precisely reflected by the graph.
However, we find that without a good way of estimating $N_{samples}$
Eq.~\ref{eq:sigma_gauss} is no longer a clearly good fit; we add an
overall scaling term $s$ and perform nonlinear multivariate
minimization with both variables $a$ and $s$, using the bootstrapped uncertainty in the
uncertainties as the weightings of the each point in the fit. This
scaling term $s$ allows us to compensate for the unknown number of
samples, since $N_{samples}^{-1/2}$ itself is simply a scaling factor.

\begin{figure*}
\begin{tabular}{cc}
\includegraphics[width=0.50\textwidth]{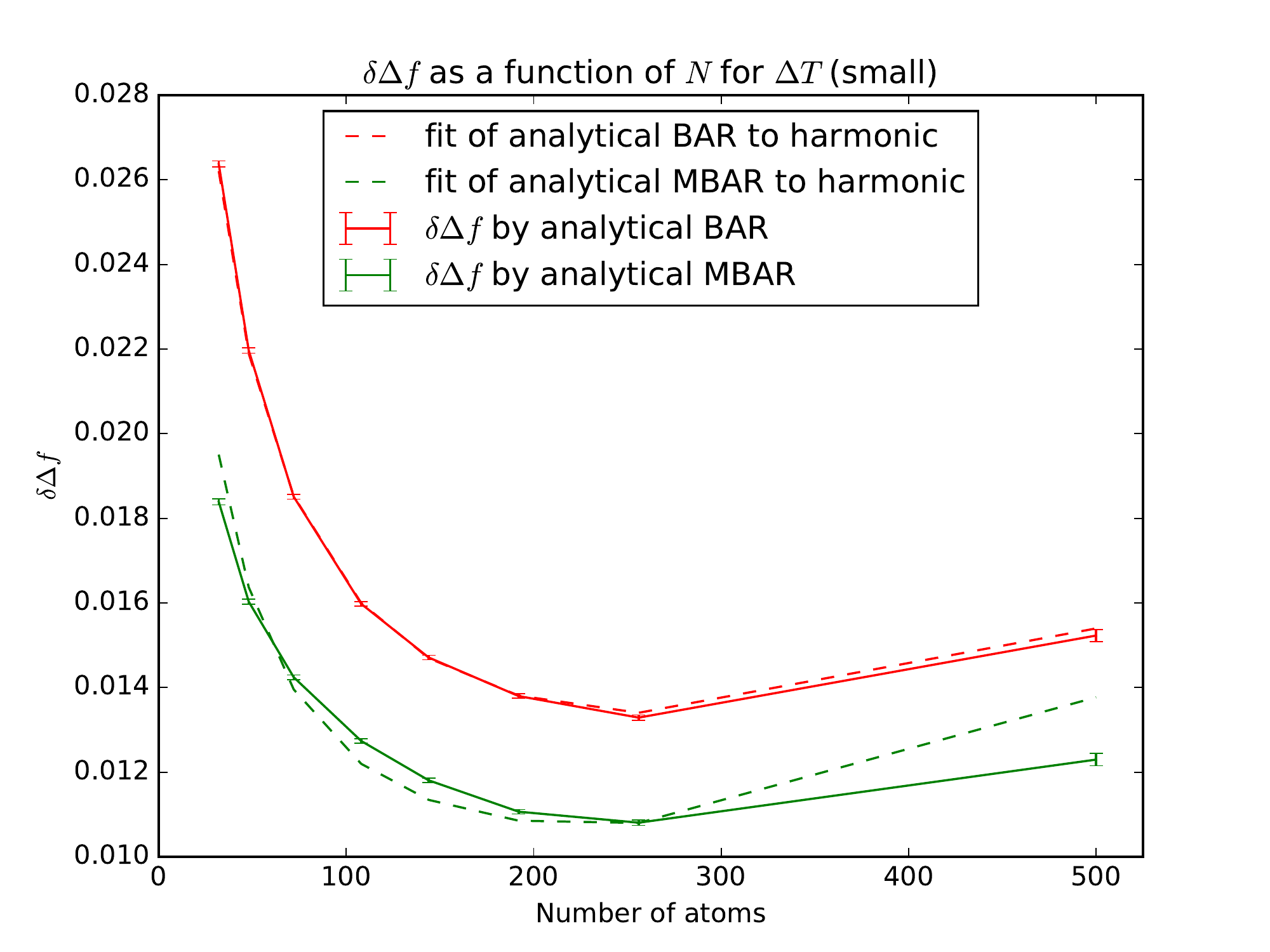} & \includegraphics[width=0.50\textwidth]{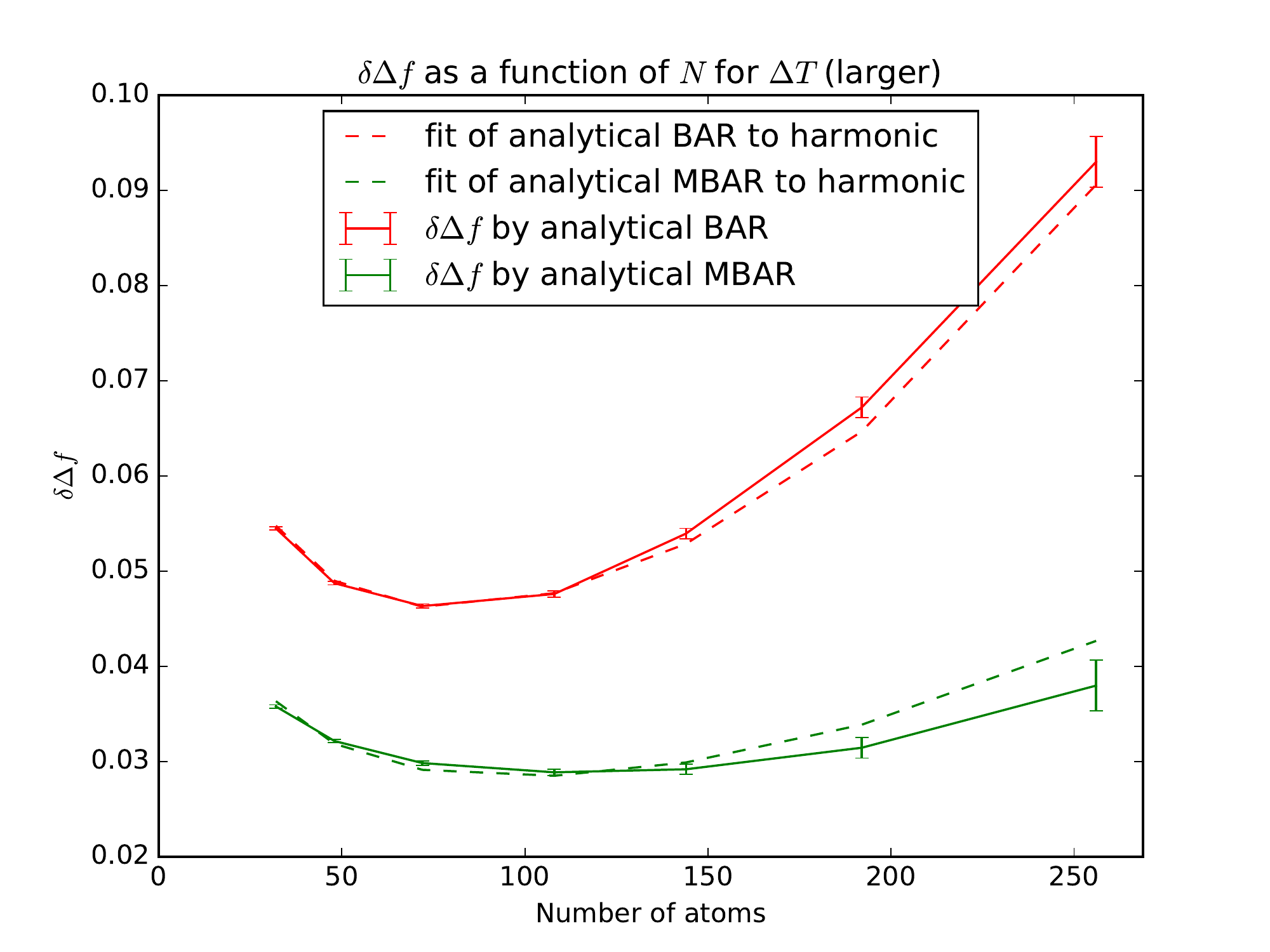}\\
\includegraphics[width=0.50\textwidth]{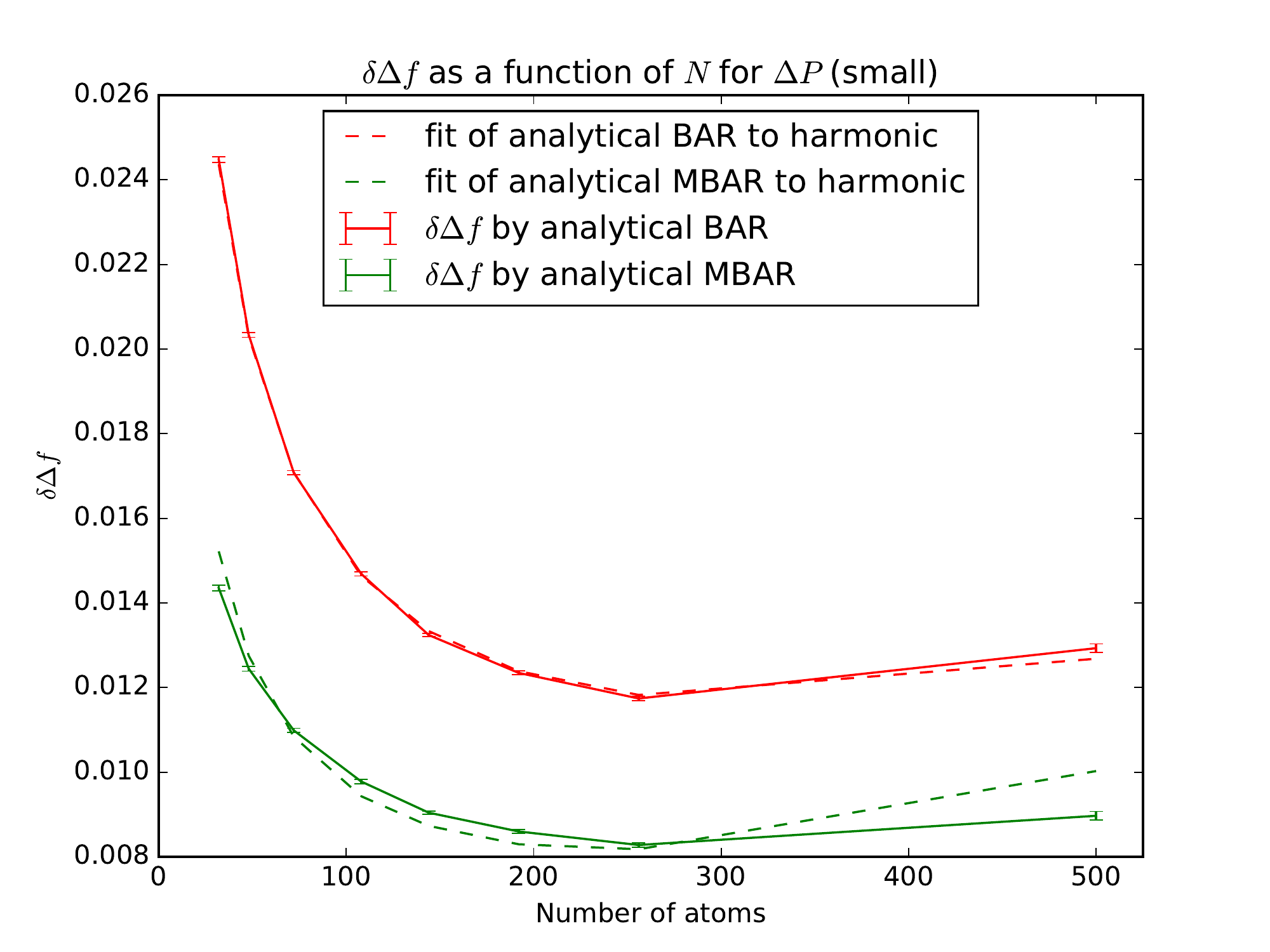} & \includegraphics[width=0.50\textwidth]{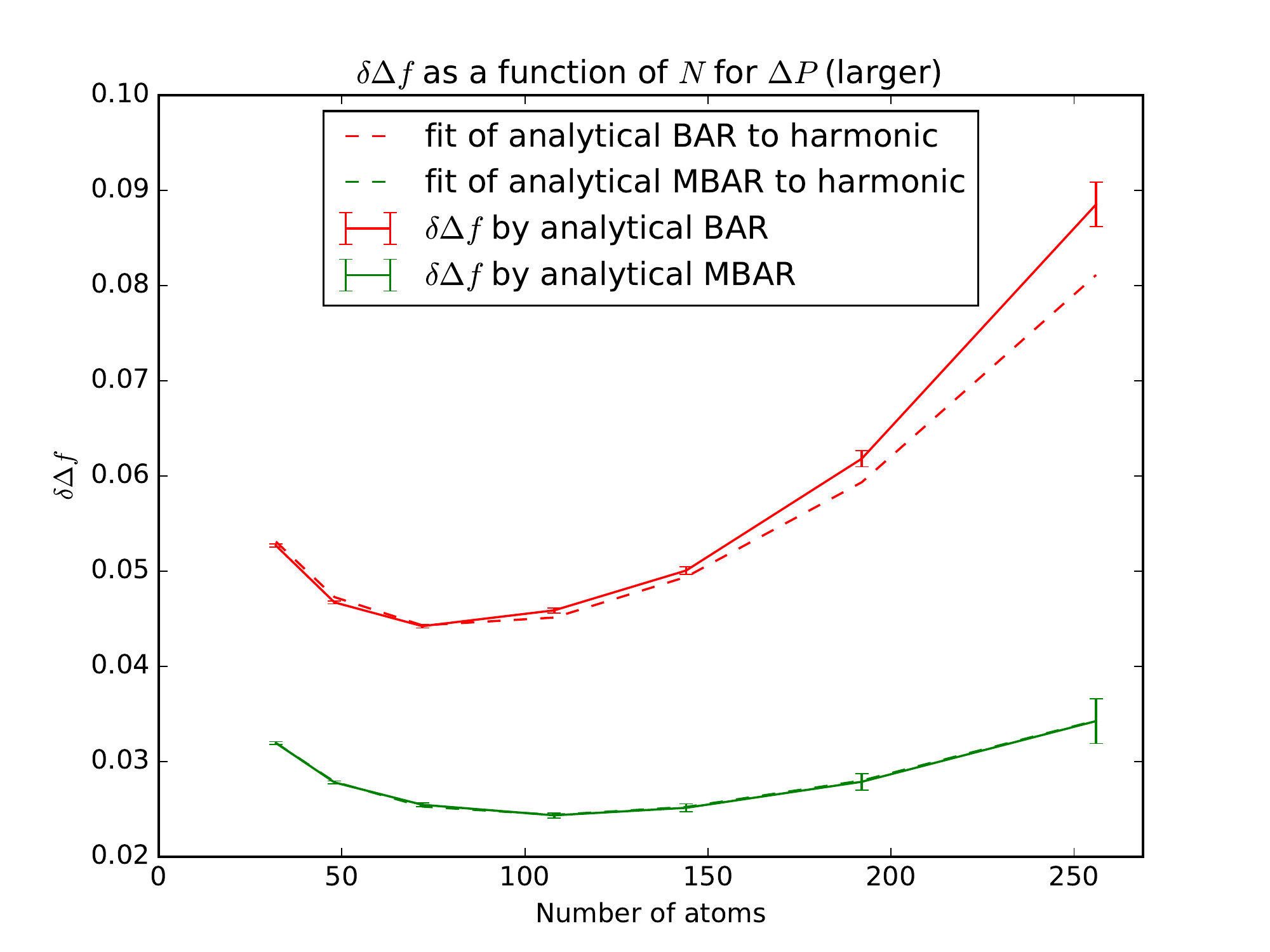}\\
\end{tabular}
\caption{The uncertainty of MBAR estimates of the reduced free energy with $\Delta T^* = 0.05$, (upper right) MBAR with $\Delta T^* = 0.1$, (lower left) MBAR with $\Delta P^* = 1.0$, (lower right) MBAR with $\Delta P^* = 2.0$, compared with the results for BAR in Fig.~\ref{fig:dTvsN} and Fig.~\ref{fig:dPvsN}.\label{fig:mbarvN}}
\end{figure*}

The results are shown in Fig.~\ref{fig:mbarvN}, compared to the results for analyzing only the two central states at a time with BAR. For clarity, we have omitted the bootstrap estimate of the variance, which is statistically indistinguishable from the analytical estimate and is somewhat noisier.

For the $\Delta T^* = 0.05$ case, the nonlinear least squares fit
approach gives $s = 0.68 \pm 0.04$, and $a = 0.048 \pm 0.004$. For
$\Delta T^* = 0.10$, $s= 0.77 \pm 0.03$ and $a = 0.112 \pm 0.005$. The
fact that the scaling factors $s$ are fairly similar indicates that
comparing $a$ is reasonable. Interestingly, $a$ increases more slowly
than quadratically with spacing, though the uncertainties involved in
this two parameter comparison make it difficult to be quantitative
rather than qualitative, However, it is clear that the minimum
uncertainty with respect to $N$ is further out than with BAR, and that
the uncertainty goes back up more slowly with $N$.

For the $\Delta P^* = 1.0$ case, the nonlinear least squares fit
approach gives $s = 0.55 \pm 0.03$, and $a = 0.045 \pm 0.003$. For
$\Delta P^* = 2.0$, $s= 0.702 \pm 0.005$ and $a = 0.105 \pm
0.001$. Scaling factors $s$ are still fairly similar, and $a$
increases more slowly than quadratically with spacing, though, again,
the uncertainties involved in this two parameter comparison.  Again,
it is clear that the minimum uncertainty with respect to $N$ is
further out than with BAR, and that the uncertainty goes back up more
slowly.

The fact that these much more complicated systems seem to follow the
behavior as simple harmonic oscillators indicates that the efficiency
scaling of system size is not as poor as originally thought. We can
actually {\em increase} the efficiency in many cases for smaller
spacings and systems.  Once we reach the size where the efficiency is the minimum,
then we can decrease the spacing to compensate, remaining roughly at
the minimum of the system.  For determination of the free energy along
a line, then the number of states to simulation to achieve a fixed
uncertainty in the phase boundary at the minimum uncertainty threshhold
will scale as $\frac{N_{new}}{N_{old}}^{1/2}$, where $N_{new}$ is the
new system size (in atoms), and $N_{old}$ is the old system size.  For
a two dimensional phase diagram, when the system size is altered, the
number of states needed to achieve the same uncertainty will then go
up by a factor of $\frac{N_{new}}{N_{old}}^{1/2}$ in each dimension,
for a factor of $(\frac{N_{new}}{N_{old}}^{1/2})^{2}$ or simply
$\frac{N_{new}}{N_{old}}$ overall. This indicates that the overall
efficiency scaling of the SIMR method goes as $N$, the size of the
system.  Since the minimum error as a function of $N$ occurs at larger $N$ for a given
spacing with MBAR, and $a$ appears to increase less than
quadratically with MBAR, it appears that MBAR scales even better with size than
BAR, though the exact behavior is harder to quantify. Therefore, given a spacing,
we can increase size until we hit the minimum in variance, as seen in
Fig.~\ref{fig:mbarvN}. As $a$ is less than 4, to first approximation,
we need to decrease spacing less as a function of size compared to BAR
(2 state reweighting) in order to stay at the variance minimum,
leading to scaling in 2D of somewhat better than $N$ and in 1D better
than $N^{1/2}$.

\section{Conclusion}

Successive interpolation of multistate reweighting (SIMR) provides an efficient and flexible method to predict polymorph phase diagrams. This method overcomes a number of the challenges in existing phase diagram prediction methods. The error does not propagate along the line and can be determined analytically with little computational expense. No previous knowledge of coexistence is required, only a reference Gibbs free energy difference  at any temperature or pressure where the phases are stable over the timescales of the simulation. This method is applicable to solid-solid coexistence, unlike the Gibbs ensemble method. A Python implementation of this method can be found at \url{http://www.github.com/shirtsgroup/phase_diagram}.

However, since the SIMR method requires sampling at states other than those directly on the coexistence line, it requires sampling at a larger number of states. The actual number of states needed is dependent on the system itself and the prior knowledge of coexistence. Also, the sampled thermodynamic states must be close enough together on the temperature-pressure plane as to have sufficient thermodynamic overlap between each set of adjacent states.

The required density of sampled states is dependent on the phase space overlap between adjacent states. The overlap between states is dependent on the number of independently moving molecules in the system and the distance between the temperatures and pressures of each state. It has been speculated that the uncertainty in the free energy difference calculations, and thus the overall efficiency, scales unfavorably with size in the regime of large numbers of molecules but more favorable within the limit of small systems, where the limit of small systems is determined by the specific system and the spacing between states. We have found that the overall scaling of the SIMR method goes as $O(N)$ where $N$ is the number of molecules in the system. 

The first full molecular dynamics solid phase diagram of crystalline benzene has been produced using this method. Three different polymorphs were simulated for the system and the reference free energy obtained from a pseudo-supercritical path was combined with multistate reweighting to generate the phase diagram. This phase diagram is qualitatively consistent with previous experimental results. The benzene phase diagram shows weak temperature dependence and strong pressure dependences, with increasing stability of polymorph II at higher pressures, consistent with experimental results.

\section{Acknowledgments}

This work used the Extreme Science and Engineering Discovery Environment (XSEDE), which is supported by National Science Foundation grant number OCI-1053575.  Specifically, it used the Bridges system, which is supported by NSF award number ACI-1445606, at the Pittsburgh Supercomputing Center (PSC). This work was supported financially by NSF through the grants NSF-CBET 1351635 and NSF-DGE 1144083. We thank Zhaoxi Sun for identifying a typo.

% LocalWords:  multistate Schieber bioavailability Lennard polymorphs Apotex al
% LocalWords:  Polymorphism metastable polymorphism Glaxo serotonin reuptake VL
% LocalWords:  GSK Rotigotine Ritonavir HIV transdermal Parkinson's tabletting
% LocalWords:  GPa Boehler Choukroun entropic Clausius Clapeyron Eike PSCP SIMR
% LocalWords:  GlaxoSmithKline paroxetine phenylbutazone Raman Duhem Dybeck QHA
% LocalWords:  fluorouracil Hof arxiv Escobedo GD Eq VA Rane subensembles Cov
% LocalWords:  binless Var Zhaoxi
% LocalWords:  Multistate SIMR multistate MBAR polymorphs pymbar NVT Frenkel al
% LocalWords:  postprocessing metadynamics Ladd PSCP intramolecular Dybeck Eq
% LocalWords:  Paliwal jacobian Lennard LJ NPS vibrational entropic GROMACS AA
% LocalWords:  anharmonicity supercell benzenes Raiteri Berendsen barostat der
% LocalWords:  Parrinello Rahman Langevin Waals integrator Nos MTTK LAMMPS XXX
% LocalWords:  Ref gridpoints kinetically Eqs
% LocalWords:  Lennard SIMR GPa polymorphs Schnieder al Raiteri DLT phonons LJ
% LocalWords:  anharmonic GROMACS Travesset pymbar MBAR pymbar Eq
% LocalWords:  SIMR Eq Ref unitless Gaussians Lennard subfigure MBAR todo scipy
% LocalWords:  pdf anaytically da phse threshhold LAMMPS refit PLACEHOLDER NPS
% LocalWords:  negligibly
% LocalWords:  multistate SIMR timescales polymorphs XSEDE OCI ACI PSC CBET DGE

%\newpage

\bibliography{SIMR}

\end{document}